%% file: VT-2023-02523.R1.tex
\documentclass[final]{IEEEtran}
\usepackage{amsthm,amssymb,graphicx,multirow,amsmath,color,amsfonts}
\usepackage[update,prepend]{epstopdf}
\usepackage[noadjust]{cite}
\usepackage[latin1, utf8]{inputenc}
\usepackage{tikz}
\usepackage{bbm} 
\usepackage{pdfpages}
\usepackage{tabulary}
\usepackage{multirow}
\usepackage{comment}
\usepackage{subfigure}
\usepackage{float}

\usepackage[titlenumbered,ruled]{algorithm2e}
\usepackage[noend]{algpseudocode}

\include{notation}

\allowdisplaybreaks 
\begin{document}
\title{Stochastic Geometry-based Trajectory Design for Multi-Purpose UAVs: Package and Data Delivery}
\author{
	Yujie Qin, Mustafa A. Kishk, {\em Member, IEEE}, and Mohamed-Slim Alouini, {\em Fellow, IEEE}
	\thanks{Yujie Qin and Mohamed-Slim Alouini are with Computer, Electrical and Mathematical Sciences and Engineering (CEMSE) Division, King Abdullah University of Science and Technology (KAUST), Thuwal, 23955-6900, Saudi 
		Arabia. Mustafa Kishk is with the Department of Electronic Engineering, Maynooth University, Maynooth, W23 F2H6, Ireland. (e-mail: yujie.qin@kaust.edu.sa; mustafa.kishk@mu.ie; slim.alouini@kaust.edu.sa).} 
	
}
\date{\today}
\maketitle
\begin{abstract}
	With the advancements achieved in drones' flexibility, low cost, and high efficiency, they obtain huge application opportunities in various industries, such as aerial delivery and future communication networks. However, the increasing transportation needs and expansion of network capacity demands for UAVs will cause aerial traffic conflicts in the future. To address this issue, in this paper, we explore the idea of multi-purpose UAVs,  which act as aerial wireless communication data relays and means of aerial transportation simultaneously to deliver data and packages at the same time. While UAVs deliver the packages from warehouses to residential areas, we design their trajectories which enable them to collect data from multiple Internet of Things (IoT) clusters and forward the collected data to terrestrial base stations (TBSs). To select the serving nearby IoT clusters, UAVs rank them based on their priorities and distances. From the perspectives of data and package delivery, respectively, we propose two algorithms that design the optimal UAVs trajectory to maximize the transmitted data or minimize the round trip time. Specifically, we use tools from stochastic geometry to model the locations of IoT clusters and TBSs. Given the nature of random locations, the proposed algorithm applies to general cases. Our numerical results show that multi-purpose UAVs are practical and have great potential to enhance the energy/time-efficiency of future networks.
\end{abstract}
\begin{IEEEkeywords}
	Stochastic geometry, multi-purpose UAVs, package delivery, data collection, IoT devices, Poisson Point Process, trajectory planning
\end{IEEEkeywords}
\section{Introduction}
As aerial transportation vehicles, unmanned aerial vehicles (UAVs), also known as drones, have attracted much attention in recent years. The technological improvements of drones make them a proper candidate as delivery vehicles. Equipped with GPS and sensors, UAVs are able to design the trajectory in real-time based on demand, detect and avoid obstacles and cross hard-to-reach areas for ground-delivery system. Improved lithium polymer batteries enable UAVs to have longer flight times and heavier payload \cite{morgan2005carbon}.
Using drones for package delivery has shown to be cost and speed-competitive compared to the traditional ground delivery method \cite{khosravi2021multi,otto2018optimization}. The potential benefits of drone delivery are low delivery costs, reduced maintenance costs, and enabling retailers to complete a customer's order to the doorstep \cite{welch2015cost}. Due to the nature of flight paths, UAVs are able to provide speedy delivery by avoiding ground obstacles and traffic jams, experiencing low aerial traffic conflicts, and hence, achieving high-speed transportation. Besides, UAVs can be used for relief delivery to disaster areas, for instance, medication or food delivery in earthquake \cite{song2018persistent}. Additionally, UAVs are expected to deliver passengers, such as drone taxis \cite{zhu2019pre}, and deliver medical products, such as blood, vaccines, and drone-based contact-less COVID-19 diagnosis and testing \cite{naren2021iomt}, which highly improve the quality of medical services.

Another important emerging application of drones is providing coverage in communication networks, such as acting as aerial base stations (BSs). Unlike traditional terrestrial base stations (TBSs), UAVs are more flexible and can quickly satisfy dynamic demands by optimizing their locations in real-time. They can adjust their positions in 3D-space to establish line-of-sight (LoS) links with ground users \cite{al2014optimal}, provide additional capacity to edge users in existing cellular networks, and establish high quality links with ground users \cite{qin2023downlink,10024838}. In addition, for the Internet of Things (IoT) devices specifically, UAV-involved networks are more suitable. Given IoT devices' limited battery capacity and transmit power, efficient communication links, e.g., high signal-to-noise (SNR) channels, are required. In this case, UAVs are considered a competitive candidate to serve these devices \cite{motlagh2016low}. UAVs can first wireless charge these devices \cite{xu2018uav,huang2019uav}, which prolongs the lifetime, and then collect data efficiently by establishing LoS links with devices and on-demand communication since these devices do not require data transmission all the time.

 Generally, UAVs are designed to be dedicated to a single purpose, which may cause heavy traffic conflicts in future networks. Motivated by the idea of building a hybrid system to provide multiple features within less space, we capture a new aspect of UAV application: multi-purpose drones. 
Specifically, we use stochastic geometry and optimization tools to design UAVs' trajectories given the random locations of IoT clusters and TBSs and analyze the system performance on the sides of package delivery and data collection/delivery.

\subsection{Related Work}
Literature related to this work can be categorized into: (i) UAV-assisted package delivery system, (ii) UAV-enabled communication networks, and (iii) stochastic geometry-based and optimization-based analysis of UAV networks. A brief discussion on related works is provided in the following lines.

{\em UAV-based package delivery system analysis.}
The speedy and cost-efficient home delivery of online goods is challenging. Using drones for last-mile delivery has received much attention. Authors in \cite{agatz2018optimization} studied the traveling salesman problem and showed that substantial savings are possible by using UAVs for delivery compared to truck-only delivery. Authors in \cite{murray2015flying} presented an integer linear programming formulations to solve a two drone delivery problems by minimizing the overall transport distance.
 Same-day delivery was analyzed in \cite{ulmer2018same} by combing drones and vehicles.
 Designing and scheduling UAVs to minimize the delivery time/distance were provided in \cite{ha2018min}.  A  creation of technology road mapping for drones, used by Amazon for their latest service Amazon Prime Air, was provided in \cite{singireddy2018technology}. Authors in \cite{yoo2018drone} examined the determinants of the customer adoption of drone delivery.  Besides packages, drones can also be used in delivering medical supplies \cite{ling2019aerial}.  Authors mentioned in \cite{8701196} that blood as well as other urgently medical supplies are delivered by Zipline to hospitals and clinics every day in Rwanda. In addition, compared with traditional ground-based delivery system, drone-based package delivery are expected to be cost-competitive and conveniently accessible in or near urban regions \cite{otto2018optimization,yoo2018drone,agatz2018optimization,poikonen2017vehicle}.

{\em UAV-based communication system analysis.}
Using drones as aerial BSs gains increasing popularity due to their high maneuverability \cite{8579209} and on-demand deployment \cite{mozaffari2019tutorial}. A survey of UAV communication networks was provided in \cite{gupta2015survey} which discussed some important issues that need to be resolved in future research areas. UAVs-assisted network in disaster areas was analyzed in \cite{9832657}. Authors in \cite{wang2019energy} use drones to assist an IoT network. By jointly optimizing the UAV trajectory and devices transmission schedule, they aimed to achieve an energy-efficient data collection  network. Single or multiple UAV platforms to collect data from IoT devices was analyzed in \cite{mozaffari2016mobile,mozaffari2017mobile}, in which TBSs are only used for backhaul. Instead of only using TBSs for backhaul links or UAV control, authors in \cite{zhan2017energy,bor2016new} jointly consider TBSs and UAVs to assist IoT networks for data collection. Authors in \cite{9754992} aimed to decrease UAV energy consumption while minimizing the task (data collection from IoT devices) completion time. A multi-UAV-enabled mobile-edge computing system was considered in \cite{8981986} where UAVs helped in offloading services for ground IoT devices. Authors in \cite{9037366} studied a UAV-enabled IoT data dissemination system and aimed to minimize the completion time by optimizing UAV trajectory and transmit power. 

{\em Stochastic geometry-based analysis of UAV networks.}  Stochastic geometry is a strong mathematical tool that enables characterizing the statistics of various aspects of large-scale  networks~\cite{7733098,6524460}, such as interference. Modeling the locations of UAVs as a Poisson point process (PPP) is widely used in literature \cite{khosravi2021multi,qin2020performance}.
By modeling the locations of TBSs and UAVs by two independent PPPs, authors in~\cite{8833522,alzenad20173} studied downlink coverage probability, average data rate and characterized the Laplace transform of the interference coming from both aerial and terrestrial BSs for the given setup.
Another commonly used point process, `Matern cluster process (MCP)', was used in~\cite{9773146,7809177,qin2021influence} to model the locations of users that exhibit a certain degree of spatial clustering and UAVs are deployed above the cluster centers to serve the cluster users. 

{\em Optimization-based analysis of UAV networks}. Authors in \cite{lyu2016placement} optimized the horizontal positions of UAVs to minimize the required number of UAVs while covering a given set of ground users. By jointly considered the altitudes and horizontal distances of UAVs, authors in \cite{bor2016efficient} optimized the 3D locations of UAVs to maximize the number of covered users. Besides static-UAV enabled networks, UAV trajectory designing and scheduling was analyzed in \cite{wu2018joint,zeng2016throughput}, in which the authors maximized the
minimum throughput of users by jointly optimizing the transmit power and UAV trajectory.  A deep-reinforcement-learning-based sparse reward scheme was proposed in \cite{wang2020deep} to address the problem of autonomous UAV navigation in large-scale complex environments. A joint UAV hovering altitude and power control optimization was studied in \cite{8489991} and the authors used Lagrange dual decomposition and concave-convex procedure method to solve the problem. To deal with the limited battery capacity issue of drones, authors in \cite{9598833} proposed a cooperative trajectory planning scheme, where a truck carrying backup batteries moves along with the UAV acting as a `mobile recharging station'.

While the existing literature mainly focus on single application of UAV-enabled network, there is few work about integrating these functions together \cite{khosravi2021multi,datapackagedelivery}. 
In our previous work \cite{datapackagedelivery}, we consider a multi-purpose UAV which deliver the package and data for a single IoT cluster simultaneously, and the authors in \cite{khosravi2021multi} consider UAVs delivering  packages while providing cellular network coverage for a certain area.  In this work, we extend our work to study the feasibility of using one UAV to serve a number of IoT clusters and deliver the package, and multiple IoT clusters system is more complex in the case of trajectory optimization, such as IoT cluster selection and traveling to TBSs.  We design two algorithms to optimize the UAV trajectories from the perspective of data delivery and package delivery, respectively, and to compare these two trajectories we define a new performance metric, data delivery efficiency, which is obtained by the ratio between collected data and the round trip time. 


\subsection{Contribution}
This paper systematically investigates the feasibility and performance of integrating different applications on a single UAV, wireless communication relays and means of transportation. While we choose IoT devices as users in the communication part of this work, it can be extended to many other components of wireless networks, such as residents, roadside units, and vehicles. Our main contributions of this work can be summarized below.

{\em Novel Framework and Performance Metrics.}
To fully explore the benefits of UAVs, we propose a novel system in which UAVs simultaneously act as aerial BSs and means of transportation. Compared with previous work, we consider UAVs simultaneously serving multiple IoT clusters with different priorities. To analyze the data delivery performance, we define a new performance metric, data delivery efficiency, which is computed by dividing the collected/delivered data by the round trip time.


{\em UAVs' Optimal Trajectory.}
Since we consider multiple IoT clusters, UAVs' trajectory optimization starts by selecting the IoT clusters and TBS(s). We propose an exhaustive search-based algorithm to order the IoT clusters based on distances and priorities, and another algorithm for decisions of traveling to TBS(s). Finally, we propose two trajectories: 
(i) for communication, we optimize the UAV trajectory to maximize the collected/delivered data for multiple IoT clusters, and (ii) for package delivery, we optimize the trajectory to minimize the round trip time while consuming all the energy to serve IoT clusters.

{\em System-Level Insights.}
Unlike existing literature, we use tools from stochastic geometry to model the locations of IoT clusters and TBSs. Since we consider all the locations to be random, the proposed optimization problems start with selecting IoT clusters based on the priorities of different types of clusters and distances. Besides, we are able to obtain the average system performance (average over locations), such as average round trip time and data delivery efficiency, under different system parameters, such as UAV battery size and delivery distance. 

\section{System Model}
\begin{figure}[ht]
	\centering
	\includegraphics[width = 1\columnwidth]{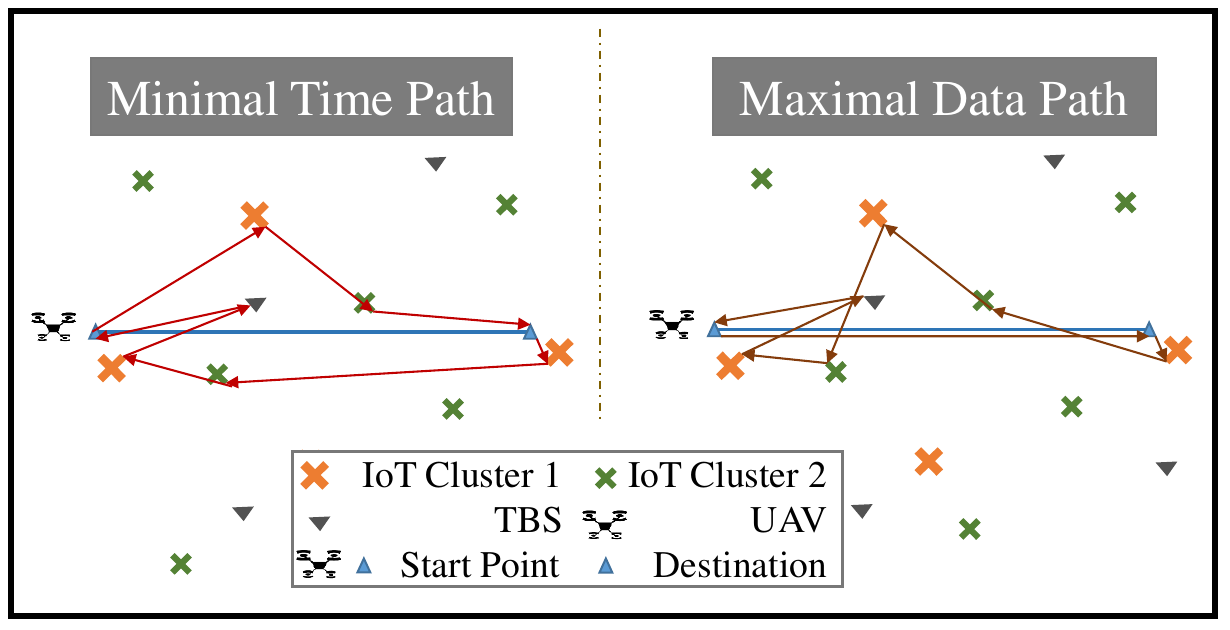}
	\caption{Illustration of the system model.}
	\label{fig_sys}
\end{figure}

We consider a multi-purpose UAV delivering a package from a warehouse to a residential area while collecting data from nearby IoT clusters and forwarding it to nearby TBSs along the route, as shown in Fig. \ref{fig_sys} and the notation used in this work is presented in Table \ref{notation}.  Let ${\rm S}$ and ${\rm D}$ denote the locations of the warehouse and residential area. Therefore, ${\rm S-D}$ pair denotes a scenario in which a UAV needs to deliver package from ${\rm S}$ to ${\rm D}$. The UAV carries a package when it starts from ${\rm S}$ and drops off the package when it arrives ${\rm D}$. Considering the priority of IoT data, say data from security monitoring and public safety is more important than the data from entertainment events,  we include two types of IoT clusters in this work. The locations of the IoT cluster centers are modeled by two independent PPPs, $\Phi_{i,1}$ and $\Phi_{i,2}$, with densities $\lambda_{i,1}$ and $\lambda_{i,2}$, respectively,   e.g., while $\Phi_{i,1}$ and $\Phi_{i,2}$ are the point processes, $x_1\in\Phi_{i,1}$ and $x_2\in\Phi_{i,2}$ denote the location of an IoT cluster center in each of the two PPPs. Assume that $\Phi_{i,1}$ has higher priority than $\Phi_{i,2}$, as well as more data required to be transferred, $M_{1} > M_{2}$,  where $M_{\{\cdot\}}$ is the size of the data which requires to be collected/delivered over the available bandwidth for simplicity of the notation. Hence the unit of $M_{\cdot}$ is bit/Hz.  The locations of TBS are modeled by another independent PPP, $\Phi_b$, with density $\lambda_t$. UAVs' trajectories are predefined, consuming all the energy to collect the data from IoT clusters and transfer all the collected data to TBSs while delivering the package. Note that we assume the  UAV hovers to collect/transmit data to achieve a more stable channel conditions. 

\begin{table}[ht]\caption{Table of Parameters}\label{notation}
	\centering
	\begin{center}
		\resizebox{1\columnwidth}{!}{
			\renewcommand{\arraystretch}{1}
			\begin{tabular}{ {c} | {c} }
				\hline
				\hline
				\textbf{Notation} & \textbf{Description} \\ \hline
				$\Phi_{b},\Phi_{i,1},\Phi_{i,2}$ & Point sets of  TBSs, type-I $\&$ type-II IoT cluter centers\\ \hline
				$\lambda_{b}$, $\lambda_{i,1}$ $\&$  $\lambda_{i,2}$	&	Density of TBSs and type-I $\&$ II IoT clusters\\ \hline
				${\rm S,D}$	&  Locations of the initial point and destination \\ \hline
				$\bar{w}$	&  Average package weight \\ \hline
				$M_1$ $\&$ $M_2$ &	Required data of type-I $\&$ II IoT clusters \\ \hline
				$r_c$	& IoT cluster radius \\ \hline
				$v_p,v_n$	 & Optimal with/without package velocity\\ \hline
				$p_{s,p},p_{s,n}$ &	Serving-related power (with/without package) \\ \hline
				$p_{m,p},p_{m,n}$ 	& Traveling-related power (with/without package)\\ \hline
				$h_u$ & 	UAV altitude\\\hline
				$B_{\rm max}$ &	Battery capacity \\\hline
				$a, b$ &	N/LoS environment variable \\\hline
				$\rho_{i},\rho_{u}$ &	Transmission power of: IoT devices, UAVs \\\hline
				$\sigma^2 $ &	Noise power\\\hline
				$\alpha_{n},\alpha_{ l}$&	N/LoS path-loss exponent \\\hline
				$m_{ n},m_{ l}$&	N/LoS fading gain  \\\hline
				$\eta_{ n},\eta_{l}$&	N/LoS additional loss
				\\\hline
				$R_{i2u}$, $R_{u2b}$ & Euclidean distances from IoT center to UAV, UAV to TBS\\\hline
				$\bar{C}_{i2u}(r)$ & Average maximum achievable rate between IoT devices to UAVs\\\hline
				$\bar{C}_{u2b}(r)$ & Average maximum achievable rate between UAVs to TBSs\\\hline		
				$T_{\cdot}$, $E_{\cdot}$ &  Time and energy-related terms\\\hline		
				$P_{\rm cov}$, $\eta$ &  Coverage probability and data delivery efficiency\\\hline\hline
		\end{tabular}}
	\end{center}
\end{table}

 UAVs serve the IoT clusters based on the distances and the priority. Without loss of generality, we select a typical ${\rm S-D}$ pair where ${\rm S}$ located at the origin and ${\rm D}$ is located at $(L,0)$. Note that in this paper we refer to the delivery UAV as `serving UAV' or `reference UAV' interchangeably. The reference UAV aims to serve $N_1$ type-I IoT clusters and use the remaining energy to serve $N_2$ type-II IoT clusters. For $i \leq N_1$, the $N_{1,i}$-th IoT cluster is the nearest to the all the possible routes among $N_{1,i-1}$ IoT clusters, ${\rm S}$ and ${\rm D}$. For $j \leq N_2$, the $N_{2,j}$-th IoT cluster is the nearest to the all the possible routes among $N_{1}$ type-I IoT clusters, $N_{2,j-1}$ type-II IoT clusters, ${\rm S}$ and ${\rm D}$. More details about the selection of the serving IoT clusters will be provided in Definition \ref{Def_IoT}. In the case of UAVs cannot collect/deliver all the required data of IoT clusters due to the limitation of the energy, it reduces the collected/delivered data from the farthest IoT clusters with the lowest priority, say the $N_{2,N_2}$-th type-II IoT clusters.

 Since all the locations are random variables and vary from realizations, we consider the reference UAVs serving the IoT clusters based on the following definition, in which the IoT clusters are selected by exhaustive search. 
\begin{definition}[Serving IoT Clusters] \label{Def_IoT}
Let $N_1$ and $N_2$ be the number of type-I and type-II IoT clusters that UAVs offer the service to. Let $\mathbf{w}_1 = \{w_{1,1},w_{1,2},...,w_{1,N_1}\}$ and $\mathbf{w}_2 = \{w_{2,1},w_{2,2},...,w_{2,N_2}\}$ be the locations of the aforementioned IoT clusters, 
\begin{align}
	w_{1,i} &= \argmin_{x_1\in\Phi_{i,1/w_{1,1},...,w_{1,i-1}}} d_1(x_1),\quad i\leq N_1,\nonumber\\
	d_1(x_1) &= \min(| x_1-y_1\times y_1|), \quad  y_1\in\{w_{1,1},...,w_{1,i-1},S,D\},\label{eq_w1_i}\\
	w_{2,j} &= \argmin_{x_2\in\Phi_{i,2/w_{2,1},...,w_{2,j-1}}} d_2(x_2),\quad j\leq N_2,\nonumber\\
	d_2(x_2) &= \min(| x_2-y_2\times y_2|), \nonumber\\ 
	& y_2\in\{ w_1,1,....,w_1,N_1,w_{2,1},...,w_{2,j-1},S,D\},\label{eq_w2_j}
\end{align}
where $y\times y$ denote the line segments formed by $y$. Consequently, the locations of the serving IoT clusters form  a two-row matrix $\mathbf{w} = \{\mathbf{w}_1,\mathbf{w}_2\}$ (since $w_{i,i}$ is a $2\times 1$ matrix denoting the location, $x$ and $y$ coordinates), which is composed of the locations of IoT clusters and starts from the cluster that has the highest priority and ends at the one has the lowest priority. 

Let $\mathbf{r} = \{r_1,r_2,...,r_{(N_1+N_2+1)!}\}$ be the set of the possible routes of the UAVs, which is a matrix containing all permutations of the elements of vector $\{\mathbf{w}_1,\mathbf{w}_2,D\}$. Each $r_i$ has $N_1+N_2+2$ elements in which $r_{i,N_1+N_2+2} = {\rm S}$  and $r_{i,1}$ to $r_{i,N_1+N_2+1}$ contains a permutation of the $(N_1+N_2+1)$ elements in $\{\mathbf{w}_1,\mathbf{w}_2,D\}$.

Conditioned on the locations of IoT clusters, for each of the possible routes, we find the locations of TBSs which are the nearest to each segment.  Let $\mathbf{w}_{b} = \{w_{b,1},w_{b,2},...,w_{b,(N_1+N_2+1)}\} $ be the locations of TBSs which UAVs may plan to go,
\begin{align}
	w_{b,i,k} &= \argmin_{x_b\in\Phi_{b}} |x_b- \overrightarrow{r_{i,k}r_{i,k+1}}|,\nonumber\\
	& r_{i,k}\in\{r_i\},1\leq k\leq N_1+N_2+1.\label{eq_wt}
\end{align}
Note that all the possible routes of UAVs should start at ${\rm S}$ and no TBS is needed from ${\rm S}$ to $r_{i,1}$ since no data collected at ${\rm S}$.
\end{definition}

\begin{figure}[ht]
	\centering
	\includegraphics[width = 1\columnwidth]{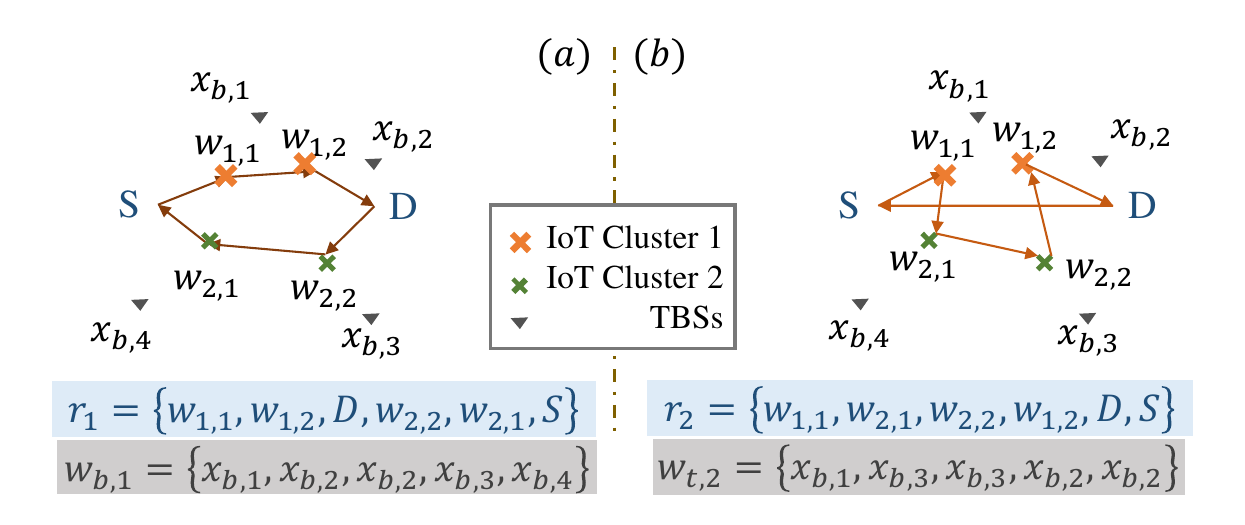}
	\caption{Illustration of the locations of the IoT clusters, possible routes and TBSs.}
	\label{fig_location}
\end{figure}
 In Fig. \ref{fig_location}, we plot two possible routes of the UAV. For given set of IoT cluster locations, the locations of TBSs can be obtained. For instance, $x_{b,1}$ is the closest TBS to the segment $\overrightarrow{w_{1,1}w_{1,2}}$ in Fig. \ref{fig_location} $(a)$. Besides, $w_{b,i}$ can contain  the repeated elements. While $\mathbf{w}$ presents the priority of IoT clusters, $\mathbf{r}$ does not contain the priority of IoT clusters.

\subsection{Power Consumption}
\label{sec_power}
UAVs rely on their internal battery for power supply, hence, the amount of flight time, payload, and transmission time are limited.
We consider the UAVs' power consumption to be composed of two parts: (i) service-related power, including hovering and communication-related power, and (ii) traveling-related power, traveling between IoT clusters, delivering packages and back to ${\rm S}$. The power consumption model of this work is based on \cite{zeng2019energy}. 

As for a rotary-wing UAV, its power consumption is sensitive to the overall payload, given by
\begin{align}
	\begin{aligned}
		p(V)=& P_{0}\left(1+\frac{3 V^{2}}{U_{\text {tip }}^{2}}\right)+P_{i}\left(\sqrt{1+\frac{V^{4}}{4 v_{0}^{4}}}-\frac{V^{2}}{2 v_{0}^{2}}\right)^{1 / 2}\nonumber\\
		&+\frac{1}{2} d_{0} \rho s A V^{3},
	\end{aligned}
\end{align}
where,
\begin{align}
	P_{0} &= \frac{\delta}{8} \rho s A \Omega^{3} R^{3},\nonumber\\
	P_{i} &= (1+k) \frac{W^{3 / 2}}{\sqrt{2 \rho A}},
\end{align}
in which $W$ is the total weight of the UAVs, $V$ is the velocity of the UAVs, $\rho$ is air density, $R$ is rotor radius, $A$ is the area of rotor disc, $v_0$ is mean rotor induced velocity, $U_{\rm tip}$ denotes the tip speed, $s$ is rotor solidity, $\Omega$ is blade angular velocity, $R$ is rotor radius, $k$ is incremental correction factor, and $\delta$ is profile drag coefficient.

Let $p_{m,n}(V)$ and $p_{s,n}(V)$ be the  motion- and service-related power of UAVs without carrying packages, and $p_{m,p}(V)$ and $p_{s,p}(V)$ be the  motion- and service-related power of UAVs while carrying packages. We consider that UAVs use the optimal velocities to minimize the energy consumption when traveling, and let $v_n$ and $v_p$ be the optimal velocities without/with package, respectively.
To simplify the notations, we consider an average weight of packages $W = W_{\rm uav}+\bar{w}_{\rm package}$. Therefore, $v_n$ and $v_p$ are constants. Consequently, we simplify the notations and use $p_{m,n}$,  $p_{m,p}$, $p_{s,n}$ and  $p_{s,p}$ since they are all constants.

\subsection{Communication Channel and Time Consumption}
To analyze the time consumption of UAVs in collecting data from IoT clusters and forwarding data  to TBSs, we first need to analyze the communication channels. The communication channels between  UAVs and (i) IoT devices (I2U), and  (ii) TBSs (U2B) are characterized by Nakagami-m fading channels. 

Given the horizontal distances between the serving UAV and IoT device, serving UAV and TBS are $R_{ i2u}$ and $R_{ u2b}$, respectively, the received power at the UAVs from IoT devices is given by
\begin{align}
	\label{eq_pi}
	&p_{i}(R_{ i2u})=\left\{\begin{array}{l}
		p_{i,l}(R_{ i2u})=\eta_{l} \rho_{i} G_{l} D_{ i2u}^{-\alpha_{l}}, \text { if } \mathrm{LoS}, \\
		p_{i,n}(R_{ i2u})=\eta_{n} \rho_{i} G_{n} D_{ i2u}^{-\alpha_{n}}, \text { if } \mathrm{NLoS},
	\end{array}\right.
\end{align}
where $D_{ i2u} = \sqrt{h_{u}^2+R_{ i2u}^2}$ and $h_{u}$ is the UAVs' altitude. Similarly, the received power at the TBSs from UAVs is given by
\begin{align}
	\label{eq_pu}
	&p_{{u}}(R_{ u2b})=\left\{\begin{array}{l}
		p_{{u,l}}(R_{ u2b})=\eta_{{l}} \rho_{{u}} G_{ l} D_{ u2b}^{-\alpha_{{l}}}, \text { if } \mathrm{LoS}, \\
		p_{{u,n}}(R_{ u2b})=\eta_{{n}} \rho_{{u}} G_{{n}} D_{ u2b}^{-\alpha_{{n}}}, \text { if } \mathrm{NLoS},
	\end{array}\right.
\end{align}
where $D_{ u2b} = \sqrt{h_{u}^2+R_{ u2b}^2}$, $\eta_{{l}}$ and $\eta_{{n}}$ are the mean additional losses for LoS and NLoS links, $\alpha_{ l}$ and $\alpha_{ n}$ are the path loss of LoS and NLoS transmissions, respectively, $G_{ l}$ and $G_{ n}$ denote the fading gains that follow Gamma distribution with shape and scale parameters $(m_{ l},\frac{1}{m_{ l}})$ and $(m_{ n},\frac{1}{m_{ n}})$, $\rho_{{i}}$ and $\rho_{{u}}$ are the transmit power of IoT devices and UAVs, respectively. The occurrence probability of LoS links and NLoS links between UAVs and serving targets (IoT devices or TBSs) are functions of Euclidean distance $r$, which are given in \cite{al2014optimal} as
\begin{align}
	P_l(r) &= \frac{1}{1+a\exp(-b(\frac{180}{\pi}\arctan(\frac{h_{u}}{r})-a))},\nonumber\\
	P_n(r) &= 1-P_l(r),\label{eq_LoSprob}
\end{align}
where $a$ and $b$ are two environment variables.

Consequently, for each of the IoT device and UAV, assuming a noise-limited communication system, the maximum achievable rate in bps/Hz is given by
\begin{align}
	C_{ i2u}(R_{ i2u}) &= \log_2\bigg(1+\frac{p_i(R_{ i2u})}{\sigma^2}\bigg).
\end{align}
Let $R_{ c2u}$ be the horizontal distances between the IoT cluster center to UAVs, the average maximum achievable rate (average over all the IoT devices within the cluster) is given by
\begin{align}
\bar{C}_{ c2u}(R_{ c2u}) \stackrel{(a)}{\approx} \log_2\bigg(1+\mathbb{E}_{R_{ i2u}}\bigg[\frac{p_i(R_{ i2u})}{\sigma^2}\bigg]\bigg),\label{eq_Cc2u}
\end{align}
where approximation in step $(a)$ follows from taking the expectation over $R_{ i2u}$ inside the logarithm operation. The reason for $\bar{C}_{\rm c2u}(R_{ c2u})$ is a function of $R_{ c2u}$ is that  IoT devices are uniformly distributed within the IoT clusters, hence, $R_{ i2u}$ is conditioned on $R_{ c2u}$ and the conditional the probability density function (PDF) is given in Lemma \ref{lemma_RRprime} (\ref{eq_Ri2u1}) and (\ref{eq_Ri2u2}). Similarly, the maximum achievable rate between UAVs and TBSs is given by
\begin{align}
	\bar{C}_{ u2b}(R_{ u2b}) = \log_2\bigg(1+\frac{p_u(R_{ u2b})}{\sigma^2}\bigg). \label{eq_Cu2b}
\end{align}

\begin{definition}[Time Consumption] \label{Def_TimeCon}
	For a certain UAV to IoT cluster and TBS link, by taking the expectation over the channel fading, given the horizontal transmission distance $R^{'} = \{R_{c2u},R_{u2b}\}$,	the transmission time of a unit data size (bit/Hz) is 
	\begin{align}
		T(R^{'}) &= \mathbb{E}_{\rm G}\bigg[\frac{1}{\bar{C}(R^{'})}\bigg] \approx \mathbb{E}_{\rm G}\bigg[\frac{1}{\log_2(1+{\rm SNR|R^{'}})}\bigg],	\label{eq_T_R}
	\end{align}
	in which ${\rm SNR|R_{\{c2u,u2b\}}} = \{\frac{p_u(R_{ u2b})}{\sigma^2},\mathbb{E}_{R_{ i2u}}[\frac{p_i(R_{ i2u})}{\sigma^2}]\}$, the subscript $G$ denotes the channel fading, and $T(R^{'})\in\{T_{ c2u}(R_{ c2u}),T_{ u2b}(R_{ u2b})\}$ corresponds to each $\bar{C}(\cdot)$ and $p(\cdot)$ as mentioned above in (\ref{eq_Cc2u}) and (\ref{eq_Cu2b}), and (\ref{eq_pi}) and (\ref{eq_pu}). 
\end{definition}

\subsection{UAV Trajectory}

 The objective of this work is to optimize UAVs' trajectory to collect/deliver data from multiple IoT clusters to TBSs while delivering packages. As mentioned, we consider that UAVs' trajectory are predefined, which enables them to forward all the collected data to TBS(s), and UAVs serve IoT clusters based on priority and distances. 

Recall that UAVs travel to nearby TBSs to forward the collected data. For a given possible routes $r_i$, UAVs may travel to multiple TBSs at any part of $r_i$ and the locations of the potential TBSs that UAVs may travel to are denoted by $\mathbf{w}_{b,i,N_1+N_2+1}$.   Let $k = 1,\cdots,N_1+N_2+1$ be the stage and $\mathbf{s} = \{s_k\}$ be the decision of each stage, where $s_{k}=0$ means that the UAV will travel from $r_{i,k}$ to the TBS $w_{b,i,k}$ then to $r_{i,k+1}$, while $s_{k}=1$ means that the UAV will travel from $r_{i,k}$ to $r_{i,k+1}$ directly.  Consequently, the modified routes of UAVs is $r^{'}_{i}\in\mathbf{r^{'}}$, which includes TBS(s). 

Note that UAVs are not necessarily hovering exactly above the IoT cluster centers and TBSs to provide service.  Instead, they can hover at a nearby point which minimizes the overall energy or time consumption. From (\ref{eq_T_R}), the transmission time decrease with the decrease of transmission distance, however, the traveling time increases. Hence, an optimal hovering point exists to minimize the overall time consumption. The same applies to energy consumption. 

\begin{definition}[Optimal Hovering Point]\label{Def_Hopt}
	For a given $r^{'}_{i}$, let $h_{i,t}$ and $h_{i,e}$ be the optimal hovering points that minimize the overall time  and energy consumption, respectively. Both $h_{i,t}$ and $h_{i,e}$ have the same length as $r^{'}_{i}$.
\end{definition}

 To simplify the notation in the following equations, we use $\mathbf{h} = \{h_1,h_2,\cdots,h_{(N_1+N_2+1)!}\}$ to represent $\mathbf{h_t} = \{h_{1,t},\cdots,h_{(N_1+N_2+1)!,t}\}$ and $\mathbf{h_e}= \{h_{1,e},\cdots,h_{(N_1+N_2+1)!,e}\}$ and $h_{1,l}$ denotes the the $l$-th point in trajectory $1$. For each of the UAV trajectory, the total time, energy consumption and collected/delivered data are given by
 \begin{align}
 T_{\rm total} &= T_{\rm col}+T_{\rm del}+T_{\rm tra},\nonumber\\
 E_{\rm total} &= E_{\rm col}+E_{\rm del}+E_{\rm tra},\nonumber\\
 M_{\rm total} &= \sum_{i = 1}^{N_1} M_{1,i}^{'} +\sum_{j = 1}^{N_2} M_{2,j}^{'},\label{eq_ttotal}
 \end{align}
where $T_{\rm col}$ and $E_{\rm col}$ denote the time and energy consumed during collecting data, $T_{\rm del}$ and $E_{\rm del}$  denote the time and energy consumed during delivering data, $T_{\rm tra}$ and $E_{\rm tra}$ are the time and energy consumed during traveling, and $M_{1,i}^{'}$ and $M_{2,j}^{'}$ are sizes of the collected/delivered data from IoT clusters, where $ M_{1,i}^{'}\leq M_1$ and $ M_{2,j}^{'}\leq M_2$,
\begin{align}
T_{\rm col} &= \sum_{i = 1}^{N_1}M_{1,i}^{'}T_{c2u}(R_{c2u,i})+\sum_{j = 1}^{N_2}M_{2,j}^{'}T_{c2u}(R_{c2u,j}),\nonumber\\
T_{\rm del} &= \sum_{k = 1}^{N_1+N_2+1}(1-s_{k})M_{k}^{''}T_{u2b}(R_{u2b,k}),\nonumber\\
T_{\rm tra} &= \sum_{l = 1}\frac{||\overrightarrow{h_{\cdot,l-1}h_{\cdot,l}}||}{v_l},\nonumber\\
E_{\rm col} &= \sum_{i = 1}^{N_1}M_{1,i}^{'}T_{c2u}(R_{c2u,i})p_{s,i}+\sum_{j = 1}^{N_2}M_{2,j}^{'}T_{c2u}(R_{c2u,j})p_{s,j},\nonumber\\
E_{\rm del} &= \sum_{k = 1}^{N_1+N_2+1}(1-s_{k})M_{k}^{''}T_{u2b}(R_{u2b,k})p_{s,k},\nonumber\\
E_{\rm tra} &= \sum_{l = 1}\frac{||\overrightarrow{h_{\cdot,l-1}h_{\cdot,l}}||}{v_l}p_{m,l},
\end{align}
in which $p_{s,k}, p_{s,i}, p_{s,j}\in\{p_{s,n}, p_{s,p}\}$, $v_l \in\{v_{n}, v_{p}\}$, $p_{m,l} \in\{p_{m,n}, p_{m,p}\}$  depend on whether the UAV is carrying the package or not, $h_{\cdot,0} = {\rm S}$ denotes the starting point of each trajectory and  $M_{k}^{''}$ denotes the delivered data for each stage and $\sum_{k}^{N_1+N_2+1}(1-s_{k})M_{k}^{''} = \sum_{i = 1}^{N_1} M_{1,i}^{'} +\sum_{j = 1}^{N_2} M_{2,j}^{'}$ which implies that UAVs delivered all the data they collected from IoT clusters.

In this work, we propose two optimization problems: (i) minimal time path, (ii) maximal data path, which optimizes UAVs' trajectories based on the round trip time and transmitted data size, respectively. For both paths, UAVs consume all the energy to collect data as long as package delivery and successful return to starting point are ensured.
\begin{definition}[Minimal Time Path and Maximal Data Path] \label{Def_MinTime}
	From the perspective of package delivery, minimal time path enables finishing a round trip quickly and deliver more packages. Let $T_{\rm total}^{*}$ be the minimal time of finishing a round trip given the number of IoT clusters required to be served. For a given realization, $T_{\rm total}^{*}$ is
\begin{align}
	\label{eq_opt_Tmin}
	T_{\rm total|\Phi_{i,1},\Phi_{i,2},\Phi_{b}}^{*} &= \min_{\mathbf{h_t},\mathbf{s}} T_{\rm total},\nonumber\\
	{\rm s.t.} &\quad  E_{\rm total} \leq B_{\rm max},\nonumber\\
	           &\quad  s_k \in \{0,1\}.
\end{align}

	From the perspective of data delivery, maximal data path enables collecting/delivering more data while delivering the package. Let $M_{\rm total}^{*}$ be the maximal transferred data while consuming all the energy. For a given realization,  $M_{\rm total}^{*}$ is
\begin{align}
	\label{eq_opt_Mmax}
	M_{\rm total|\Phi_{i,1},\Phi_{i,2},\Phi_{b}}^{*}& = \max_{\mathbf{h_e},\mathbf{s}} M_{\rm total},\nonumber\\
	{\rm s.t.} &\quad  E_{\rm total} \leq B_{\rm max},\nonumber\\
	&\quad  s_k \in \{0,1\}.
\end{align}

For both optimization problems, $\mathbf{r}$ and $\mathbf{s}$ are the possible UAV routes and decisions to travel to TBS(s), respectively.
\end{definition}

Notice that the above optimization problems solve for conditional realizations, conditioned on the realizations of $\Phi_{i,1},\Phi_{i,2},\Phi_{b}$, and we are interested in general performance. Besides, to better investigate the data delivery efficiency, we define \textit{data delivery efficiency} $\xi$ to characterize the collected/delivered data per round trip time.
\begin{definition}[Data Delivery Efficiency] \label{Def_dataeff}
 Data delivery efficiency, which characterizes the system average data collection performance (average over the locations), is defined as
\begin{align}
\xi = \mathbb{E}_{\Phi_{i,1},\Phi_{i,2},\Phi_{b}}\bigg[\frac{M_{\rm total|\Phi_{i,1},\Phi_{i,2},\Phi_{b}}^{*'}}{T_{\rm total|\Phi_{i,1},\Phi_{i,2},\Phi_{b}}^{*'}}\bigg],
\end{align} 
in which $M_{\rm total|\Phi_{i,1},\Phi_{i,2},\Phi_{b}}^{*'}$ and $T_{\rm total|\Phi_{i,1},\Phi_{i,2},\Phi_{b}}^{*'}$ are the transmitted data and round trip time of the optimal trajectory.
\end{definition}
The higher $\xi$, the higher system energy efficiency, and this performance metric is upper-bounded by the average of the maximum achievable data rates of I2U and U2B channels.

 In what follows, we start the performance analysis of this work. The structure and relations between the definitions, lemmas, and theorems are summarized as follows. The goal of this work  (Definition 4) is obtaining the optimal trajectory of package delivery, minimal time path, and optimal trajectory of data delivery, maximum data path. Since the locations of IoT clusters and TBSs are all random we first need to select the IoT clusters to serve and TBS(s) to communicate (Definition 1, and analysis is provided in Algorithm 1). The data collection time and energy consumption, which are required for trajectory design, are defined in Definition 2 and analyzed in Lemma 2 and Theorem 1. Besides, we define the trade-off between communication time (energy) and traveling time (energy) in Definition 3 and analyze it in  Lemma 3 and Lemma 4, and study the time (energy)-efficiency of traveling to TBS(s) in Lemma 5. Finally, we propose Algorithm 4 to obtain the final trajectory of the UAV. 

\section{Performance Analysis}

This section aims to analyze the locations and the energy and time consumption of data delivery. To do so, we first propose an algorithm to select IoT clusters and then analyze the data rate given that the locations of IoT clusters and TBSs are randomly distributed.

\subsection{Locations and Distance Analysis}

Before analyzing the communication time consumption, we first need to obtain the locations of IoT clusters. Recall that UAVs serve the IoT clusters based on distances and priorities.
  Take $N_1 = 2$ and $N_2 = 2$ for example, given in Definition \ref{Def_IoT}, the location matrix of serving IoT clusters is $\mathbf{w} = \{w_{1,1},w_{1,2},w_{2,1},w_{2,2}\}$. The reference UAV checks if it is able to collect/deliver all the required data starting from $w_{1,1}$ to $w_{2,2}$. If it is able to deliver data for $w_{1,1}$ then check $w_{1,2}$. If not, deliver part of the data for $w_{1,1}$ based on minimal time or maximal data path policies.

We propose the Algorithm \ref{Alg_iotlocation} to order the IoT clusters given a realization of the locations $\Phi_{i,1},\Phi_{i,2},\Phi_{b}$.

	\begin{algorithm}
		\caption{Algorithm for Serving IoT Cluster Locations}
		\SetAlgoLined
		\DontPrintSemicolon
		\KwIn{$N_1,N_2$: Number of serving IoT clusters\newline
			$x_1,x_2$: Set of locations of two types of IoT cluster centers\newline
			$S,D$: Locations of the source and destination}
		\KwOut{$\mathbf{w} = \{w_1,w_2\}$: Locations/orders of serving IoT clusters }
		\textbf{Initialization:} $\mathbf{w} = \emptyset$, $w_{1,0} = \emptyset$, $w_{2,0} = \emptyset$, $i = j =1$, $w^{'} = \{S,D\}$ \newline
		\SetKwFunction{FMain}{IoTCluster}
		\SetKwProg{Fn}{Function}{:}{}
		\Fn{\FMain{$N_1,N_2,x_1,x_2$}}{
		\ForEach{$ i \leq N_1$}{
	     Solve (\ref{eq_w1_i}) for given $w_{1,i-1}$ and $w^{'}$, and denote the solution as $w_{1,i}$.
			}
		\ForEach{$ j \leq N_2 $}{
			Solve (\ref{eq_w2_j}) for given $w_{2,j-1}$, $w^{'}$, and $w_{1}$, and denote the solution as $w_{2,j}$.
		}
			\textbf{return} ${\mathbf{w} = \{w_1,w_2\}}$ 
		}
		\textbf{End Function}
		\label{Alg_iotlocation}
	\end{algorithm}
The locations of TBSs follow a same method, thus omitted here.

To analyze the transmission time between UAVs and IoT devices, we need to obtain the distance distribution of $R_{i2u}$, which is conditioned on $R_{c2u}$. As mentioned in Definition \ref{Def_Hopt}, UAVs hover at the optimal points to communicate with TBSs and IoT clusters. Hence, $R_{c2u}$ and $R_{u2b}$ are not random variables, instead, they are predefined for the given route.
\begin{lemma}[Distribution of $R_{ i2u}$]\label{lemma_RRprime}
	Given the distance between the IoT cluster center and the serving UAV is $R_{\rm c2u}$, in the case of $R_{ c2u}>r_c$, where $r_c$ is the radius of IoT devices cluster, the PDF  of $R_{ i2u}$ is given by
	\begin{align}
		\label{eq_Ri2u1}
		f_{ R_{i2u}}(r) &= \frac{2r}{\pi r_c^2}\arccos\bigg(\frac{R_{ c2u}^2+r^2-r_c^2}{2 R_{ c2u} r}\bigg), \nonumber\\
		& (R_{ c2u}-r_c<r<R_{ c2u}+r_c),
	\end{align}
	otherwise, if $R_{\rm c2u}\leq r_c$, the PDF of $R_{ i2u}$ is
	\begin{align}
		\label{eq_Ri2u2}
		f_{ R_{i2u}}(r) &= \left\{\begin{aligned}
			&\frac{2r}{r_c^2},\quad 0<r<r_c-R_{\rm c2u} \\
			&\frac{2r}{\pi r_c^2}\arccos\bigg(\frac{R_{ c2u}^2+r^2-r_c^2}{2 R_{ c2u} r}\bigg),\\ &\qquad r_c-R_{ c2u}<r<R_{ c2u}+r_c,
		\end{aligned}\right.
	\end{align}
 and for other $r$, $f_{R_{i2u}}(r) = 0$.
\end{lemma}

\subsection{Time Consumption Analysis}
Obtaining the conditional PDF of $R_{ i2u}$, we are able to compute coverage probabilities (CCDF of SNR) for both IoT to UAV and UAV to TBS links as functions of $R_{ c2u}$ and  $R_{ u2b}$, respectively.
Observe that the time consumption defined in Definition \ref{Def_TimeCon} requires to take the expectation over SNR. Using the following lemma, we derive the PDF of SNR.
\begin{lemma}[PDF of SNR]
	Coverage probability is the CCDF of SNR, given by
		\begin{align}
			P_{\rm cov|R_{c2u}} &= \mathbb{P}\bigg(\frac{p_{i}(R_{i2u})}{\sigma^2}>\gamma\bigg) \nonumber\\
			&=  \int_{r} \bigg(\sum_{k = 0}^{m_l}\frac{(m_l g_l(\sqrt{r^2+h_{u}^2})\gamma)^k}{k!}P_l(\sqrt{r^2+h_{u}^2})\nonumber\\
			&\quad\times\exp(-m_l g_l(\sqrt{r^2+h_{u}^2}) \gamma)\nonumber\\
			&+\sum_{k = 0}^{m_n}\frac{(m_n g_n(\sqrt{r^2+h_{u}^2})\gamma)^k}{k!}P_n(\sqrt{r^2+h_{u}^2})\nonumber\\
			&\quad\times\exp(-m_n g_n(\sqrt{r^2+h_{u}^2}) \gamma)\bigg)f_{\rm R_{i2u}}(r) {\rm d}r,\nonumber\\
			P_{\rm cov|R_{u2b}} &= \mathbb{P}\bigg(\frac{p_{u}(R_{ u2b})}{\sigma^2}>\gamma\bigg) \nonumber\\
			&=   \bigg(\sum_{k = 0}^{m_l}\frac{(m_l g_l(\sqrt{R_{ u2b}^2+h_{u}^2})\gamma)^k}{k!}P_l(\sqrt{R_{ u2b}^2+h_{u}^2})\nonumber\\
			&\quad\times\exp(-m_l g_l(\sqrt{R_{ u2b}^2+h_{u}^2}) \gamma)\nonumber\\
			&+\sum_{k = 0}^{m_n}\frac{(m_n g_n(\sqrt{R_{ u2b}^2+h_{u}^2})\gamma)^k}{k!}P_n(\sqrt{R_{ u2b}^2+h_{u}^2})\nonumber\\
			&\quad\times\exp(-m_n g_n(\sqrt{R_{ u2b}^2+h_{u}^2}) \gamma)\bigg),
		\end{align}
	where $g_l(r) =  \gamma (\rho\eta_{{l}})^{-1} r^{\alpha_{{l}}}$ and $g_n(r) = \gamma (\rho\eta_{{n}})^{-1} r^{\alpha_{{n}}}$.
	Hence, the PDF of SNR is derived by taking the first derivative of CCDF,
		\begin{small}
			\begin{align}
			&f_{\rm SNR|R_{c2u}}(\gamma) 
			= \sum_{k = 1}^{m_l-1}\int_{r}\frac{(m_l g_l(\sqrt{r^2+h_{u}^2}))^k}{k!}P_l(\sqrt{r^2+h_{u}^2})\nonumber\\
			&\exp(-m_l g_l(\sqrt{r^2+h_{u}^2}) \gamma)f_{\rm R_{c2u}}(r)(m_l g_l(\sqrt{r^2+h_{u}^2})\gamma^{k}-k\gamma^{k-1}) {\rm d}r\nonumber\\
			&+\sum_{k = 1}^{m_n-1}\int_{r}\frac{(m_n g_n(\sqrt{r^2+h_{u}^2}))^k}{k!}P_n(\sqrt{r^2+h_{u}^2})\nonumber\\
			&\quad\times\exp(-m_n g_n(\sqrt{r^2+h_{u}^2}) \gamma)f_{ R_{i2u}}(r)\nonumber\\
			&\quad\times(m_n g_n(\sqrt{r^2+h_{u}^2})\gamma^{k}-k\gamma^{k-1}) {\rm d}r,\nonumber\\
			& f_{\rm SNR|R_{u2b}}(\gamma) 
			= \sum_{k = 1}^{m_l-1}\frac{(m_l g_l(\sqrt{R_{ u2b}^2+h_{u}^2}))^k}{k!}P_l(\sqrt{R_{ u2b}^2+h_{u}^2})\nonumber\\
			&\times\exp(-m_l g_l(\sqrt{r^2+h_{u}^2}) \gamma)(m_l g_l(\sqrt{R_{ u2b}^2+h_{u}^2})\gamma^{k}-k\gamma^{k-1}) \nonumber\\
			&+\sum_{k = 1}^{m_n-1}\frac{(m_n g_n(\sqrt{R_{ u2b}^2+h_{u}^2}))^k}{k!}P_n(\sqrt{R_{ u2b}^2+h_{u}^2})\nonumber\\
			&\exp(-m_n g_n(\sqrt{R_{ u2b}^2+h_{u}^2}) \gamma)(m_n g_n(\sqrt{R_{ u2b}^2+h_{u}^2})\gamma^{k}-k\gamma^{k-1}).
		\end{align}
		\end{small}
\end{lemma}
\begin{IEEEproof}
	The coverage probability equations is derived by
\begin{small}
		\begin{align}
	&P_{\rm cov|R_{c2u}} = \mathbb{P}\bigg(\frac{p_{i}(R_{i2u})}{\sigma^2}>\gamma\bigg)= P_{n}(R_{i2u})\mathbb{P}\bigg(\eta_{l}\rho_i G_l D_{i2u}^{-\alpha_l}>\gamma\sigma^2\bigg) \nonumber\\
	&\quad+P_{l}(R_{i2u})\mathbb{P}\bigg(\eta_{n}\rho_i G_n D_{i2u}^{-\alpha_n}>\gamma\sigma^2\bigg)\nonumber\\
	&= P_{n}(R_{i2u})\mathbb{P}\bigg( G_l >\frac{D_{i2u}^{\alpha_l}\gamma\sigma^2}{\eta_{l}\rho_i}\bigg)+P_{l}(R_{i2u})\mathbb{P}\bigg( G_n >\frac{D_{i2u}^{\alpha_n}\gamma\sigma^2}{\eta_{n}\rho_i}\bigg),
	\end{align}
\end{small}
the proof completes by notice that (i)  $\Bar{F}_{\rm G}(g)=\frac{\Gamma_{u}(m,g)}{\Gamma(m)}$, where $\Gamma_{u}(m,g)=\int^{\infty}_{mg}t^{m-1}e^{-t}{\rm d}t$ is the upper incomplete Gamma function, and (ii)  $\frac{\Gamma_{u}(m,g)}{\Gamma(m)}=\exp(-g)\sum^{m-1}_{k=0}\frac{g^{k}}{k!}$.
\end{IEEEproof}

\begin{theorem}[Time Consumption] Following Definition \ref{Def_TimeCon}, we take the expectation over SNR using the PDF derived above, the time consumption given the horizontal distance is 
	\begin{align}
		&T_{\{c2u,u2b\}}(\{R_{c2u},R_{u2b}\}) \nonumber\\
		&= \int_{0}^{\infty}\frac{1}{\log_2(1+{\rm SNR|\{R_{c2u},R_{u2b}\}})}f_{\rm SNR|\{R_{c2u},R_{u2b}\}}(\gamma) {\rm d}\gamma,\label{eq_T(R)}
	\end{align}
 and the unit of the transmission time is sec$\times$Hz/bit.  
 
Recall that this term $	T_{\{c2u,u2b\}}(\{R_{c2u},R_{u2b}\})$ is multiplied by $M$ to compute the total travel time, and $M$ is defined as the ratio between the data size and the available bandwidth (in bit/Hz), hence, the product will be in seconds.
\end{theorem}
We have obtained the locations of IoT clusters and TBSs in Algorithm \ref{Alg_iotlocation}, and communication time of transmitting unit data (\ref{eq_T(R)}). We are able to proceed to UAV trajectory optimization.

\section{UAV Trajectory Optimization}

In this section, we propose the final algorithms of obtaining UAVs' minimal time and maximal data path. The algorithms are composed of (i) the IoT cluster selection and communication analysis from the previous section, and (ii)  finding optimal hovering points and decisions to traveling to TBSs in the following section. More details are provided below.

\subsection{Optimal Hovering Point}
Recall that UAVs are not necessarily to travel to exactly above the IoT cluster centers and TBSs, instead, they can hovering at a nearby point to communicate. 

We first solve this optimal hovering point problem in the case of a single IoT cluster/TBS.
Suppose now the reference UAV travels from $A$ to $B$ while hovering at the location $h(d)$ to communicate with $c$ at $d$ away, where $c$ can be either a TBS or an IoT cluster center. The energy and time of the given path are
\begin{align}
E_{h} &= T_{\{\cdot\}}(d)M_t p_{s,c}+(|A-h(d)|+|B-h(d)|)\frac{p_{m,c}}{v_{c}},\label{eq_Eh}\\
T_{h} &= T_{\{\cdot\}}(d)M_t+(|A-h(d)|+|B-h(d)|)\frac{1}{v_{c}},\label{eq_Th}
\end{align}
where  $T_{\{\cdot\}}(d)$ is defined in (\ref{eq_T(R)}), $M_t$ is the transmitted data, $p_{s,c}\in \{p_{s,n},p_{s,p}\}$, $p_{m,c}\in \{p_{m,n},p_{m,p}\}$ and $v_{c}\in \{v_{n},v_{p}\}$ which depend on whether the UAV is carrying the package or not.

\begin{figure}[ht]
	\centering
	\includegraphics[width = 1.1\columnwidth]{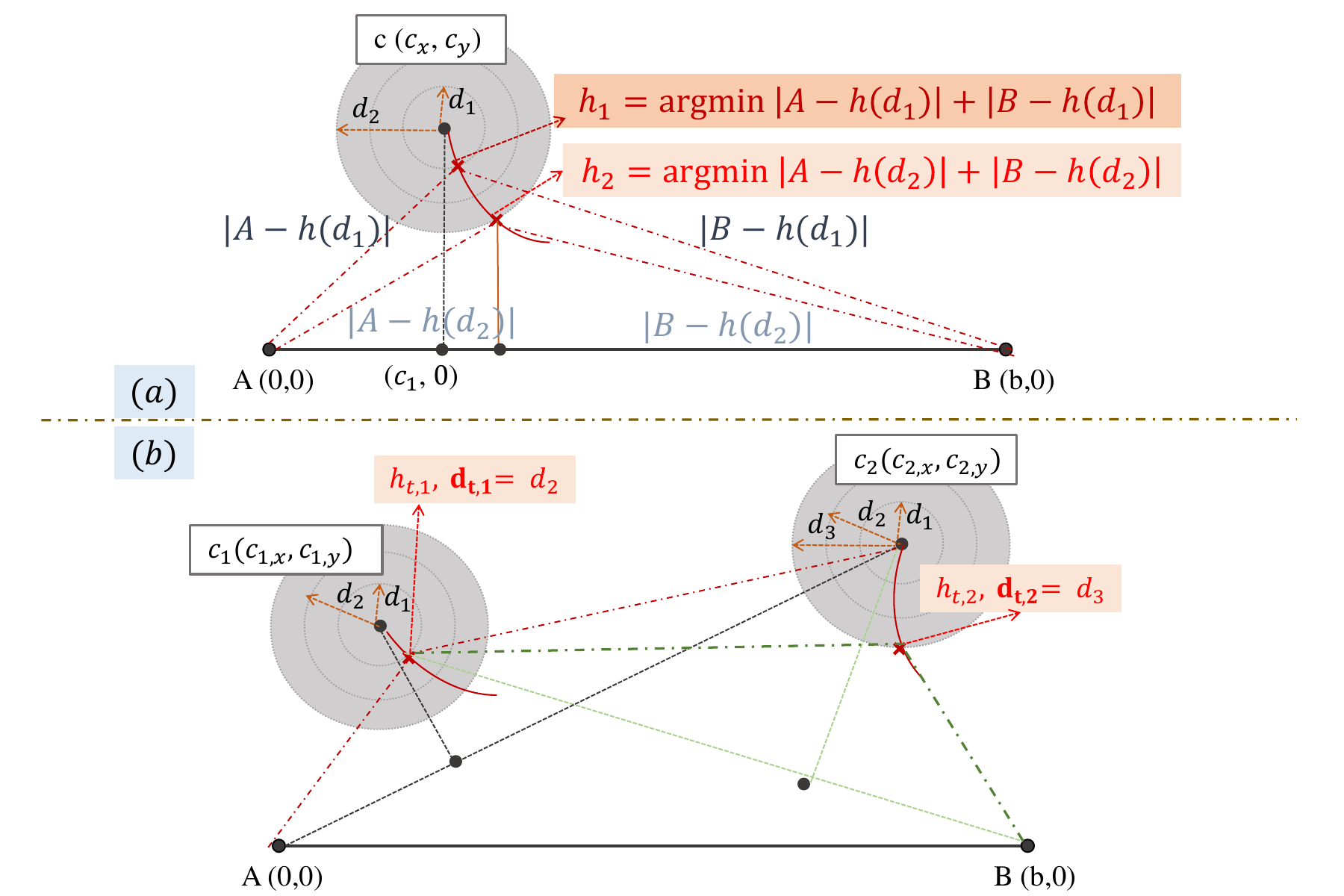}
	\caption{Illustration of the optimal hovering point for \textbf{(a)} a single user,  \textbf{(b)} multiple users.}
	\label{fig_hoverpoint}
\end{figure}

\begin{lemma}[Optimal Hovering Point]\label{lemma_hovering}
	(\ref{eq_Eh}) and 	(\ref{eq_Th}) can be solved numerically since taking the derivative of $T_{\{\cdot\}}(d)$ over $d$ is difficult to compute. We observe that $E_{h}$ and $T_{h}$ are only functions of the path length $|A-h(d)|+|B-h(d)|$ for a given $d$,
	\begin{align}
		l^{*}(d) &= \min_{h(d)\in\R^{2}} |A-h(d)|+|B-h(d)|,\nonumber\\
		h &= \argmin_{h(d)\in\R^{2}} |A-h(d)|+|B-h(d)|,\nonumber
	\end{align}
as shown in Fig. \ref{fig_hoverpoint} $(a)$. Consequently, the solutions to (\ref{eq_Eh}) and 	(\ref{eq_Th}) are given by
\begin{align}
	h_e& = \argmin_{h(d_e)\in\R^{2}} |A-h(d_e)|+|B-h(d_e)|, \nonumber\\
	h_t &= \argmin_{h(d_t)\in\R^{2}} |A-h(d_t)|+|B-h(d_t)|,\nonumber
\end{align}
where,
\begin{align}
		d_e &= \argmin_{d\in(0,d_{\rm max})} T_{\{\cdot\}}(d)M_t p_{s,c}+l^{*}(d)\frac{p_{m,c}}{v_{c}},\nonumber\\
		d_t &= \argmin_{d\in(0,d_{\rm max})} T_{\{\cdot\}}(d)M_t+l^{*}(d)\frac{1}{v_{c}},\nonumber
	\end{align}
where $d_{\rm max}$ is the distance between $c$ to ${\rm A-B}$.
\end{lemma}

Now we extend the above results to multiple IoT clusters/TBSs scenario, replacing $c$ by $\mathbf{c}$, and corresponding optimal hovering points are denoted by $\mathbf{h}$,
\begin{align}
	E_{\mathbf{h}} &= T_{\{\cdot\}}(\mathbf{d})\mathbf{M_t} p_{s,\mathbf{c}}+(|A-\mathbf{h_1(d_1)}|+|B-\mathbf{h_n(d_n)}|\nonumber\\
	&+\sum_{i = 2}^{n}|\mathbf{h_i(d_i)}-\mathbf{h_{i-1}(d_{i-1})}|)\frac{p_{m,\mathbf{c}}}{v_{\mathbf{c}}},\\
	T_{\mathbf{h}} &= T_{\{\cdot\}}(\mathbf{d})\mathbf{M_t}+(|A-\mathbf{h_1(d_1)}|+|B-\mathbf{h_n(d_n)}|\nonumber\\
	&+\sum_{i = 2}^{n}|\mathbf{h_i(d_i)}-\mathbf{h_{i-1}(d_{i-1})}|)\frac{1}{v_{\mathbf{c}}},
\end{align}
where $p_{s,\mathbf{c}}\in \{p_{s,n},p_{s,p}\}$, $p_{m,\mathbf{c}}\in \{p_{m,n},p_{m,p}\}$ and $v_{\mathbf{c}}\in \{v_{n},v_{p}\}$ which depend on whether UAVs deliver the package or not, and $\mathbf{M_t}$ is a vector in multiple IoT clusters/TBSs scenario since the transmission data vary from each cluster/TBS.  Note that in Lemma \ref{lemma_hovering}, there exists a trade-off between transmission time (energy) and traveling time (energy) since shorter communication distances result in a better communication channel (shorter transmission distance and lower energy consumption) but a longer traveling distance (higher traveling energy consumption). Therefore, Lemma \ref{lemma_hovering} actually computes the trade-off between these two times (energy consumption). 

\begin{lemma}[Minimal Energy and Time path]\label{lemma_H1H2}
Let $\mathbf{h_t}$ and $\mathbf{h_e}$ be the hovering points which minimize overall time and energy of the trajectory of UAV given the locations of $\mathbf{c}$, 
\begin{align}
	\mathbf{h_t} &= \argmin_{\mathbf{h}\in\R^{n\times 2}} T_{\mathbf{h}},\\
	\mathbf{h_e} &= \argmin_{\mathbf{h}\in\R^{n\times 2}} E_{\mathbf{h}}.
\end{align}
As shown in Fig. \ref{fig_hoverpoint} $(b)$, we first solve the optimal hovering point for $c_1$ and then use the solution to solve for $c_2$. Solutions converge with the increase of iterations.
\end{lemma}
In what follows, we propose an algorithm to solve the above optimization problems in (\ref{eq_opt_Tmin}) and (\ref{eq_opt_Mmax}), in which $\mathbf{h} = \{h_1,h_2,\cdots\}$ is the matrix formed by the locations of hovering points and $\mathbf{d} = \{d_1,d_2,\cdots\}$ is the distance between the hovering points and IoT clusters/TBSs  locations.

		\begin{algorithm}
			\caption{Algorithm for hovering points}
			\SetAlgoLined
			\DontPrintSemicolon
			\KwIn{$S,D,\mathbf{c}$: Locations}
			\KwOut{$\mathbf{h}$: Set of locations of hovering points }
			\textbf{Initialization:} $\mathbf{d}^{0}=0,\mathbf{h}^{0}=\mathbf{r}$, $l= 0$ \newline
			\SetKwFunction{FMain}{HoverPoint}
			\SetKwProg{Fn}{Function}{:}{}
			\Fn{\FMain{$S,D,\mathbf{c}$}}{
			\SetKwBlock{Repeat}{repeat}{}
				\Repeat{
			Solve (\ref{eq_Eh}) or (\ref{eq_Th}) for $h_{1}^{l}$ given $A$  and $B = h_{2}^{l}$, and denote the optimal solution as $h_{1}^{l+1}$ and update $d_{1}^{l+1}$\\
			Solve (\ref{eq_Eh}) or (\ref{eq_Th}) for $h_{2}^{l}$ given $A = h_{1}^{l+1}$  and $B = h_{3}^{l}$, and denote the optimal solution as $h_{2}^{l+1}$ and update $d_{2}^{l+1}$\\
			Solve (\ref{eq_Eh}) or (\ref{eq_Th}) for $h_{i}^{l}$ given $A = h_{i-1}^{l+1}$  and $B = h_{i+1}^{l}$, and denote the optimal solution as $h_{i}^{l+1}$ and update $d_{i}^{l+1}$\\
		Update $l = l+1$}
				The element-wise increase of $|\mathbf{d^{l+1}}-\mathbf{d^{l}}|$ is
				below a threshold $\epsilon>0$\;
}\textbf{End Function}
		\label{Alg_path}
		\end{algorithm}

\subsection{Decisions of Traveling to TBSs}
The decisions of traveling to TBSs are the final requirement to finalize the optimal UAV trajectory.

 As mentioned, UAVs deliver the data to nearby TBSs. Recall that for each of the possible routes,  $r_i$ mentioned in Definition \ref{Def_IoT}, we have already obtained the locations of nearby TBSs, which is $w_{b,i}$ and defined $k = 1,2,3,...$ be the stage, $s_k \in \{0,1\}$ be the decisions of each stage. Now we need $s_k$ for stage $k$ and obtain the modified route $r^{'}_{i}$. To do so, we first ignore the horizontal transmission  distance between UAVs and IoT cluster centers (assuming UAVs hovering exactly above IoT cluster centers) and compute the needed energy and time without visiting TBSs, denoted by $T_{\rm noTBS}$ and $E_{\rm noTBS}$. We consider traveling to TBSs and forwarding the data as additional cost of each stage.
 
 To find the optimal TBS(s) to forward the data for a given UAV trajectory (given the locations of IoT clusters and visiting order), we solve the following optimization problem.
 
  Let $D_{c,k}$, $D_{d,k}$, $p_{s,k}$,  $p_{m,k}$ and $v_{k}$ be the states of each stage, in which $D_{c,k}$ and $D_{d,k}$ be the size of  data that the UAV currently collected (without forwarding to TBSs) and size of data required to be collected at stage $k$ and $p_{s,k}\in \{p_{s,n},p_{s,p}\}$, $p_{m,k}\in \{p_{m,n},p_{m,p}\}$ and $v_{k}\in \{v_{n},v_{p}\}$ which depend on whether UAVs delivered the package or not. Assume that UAVs forwarding all the collected data  once they connect with a TBS. 
Similarly, we here assume that $d_{u2b} = 0$, which implies that UAVs hovering exactly above the TBSs to communicate. In this way, we can obtain a solution for $s_k$.

\begin{figure*}[ht]
	\centering
	\includegraphics[width = 1.6\columnwidth]{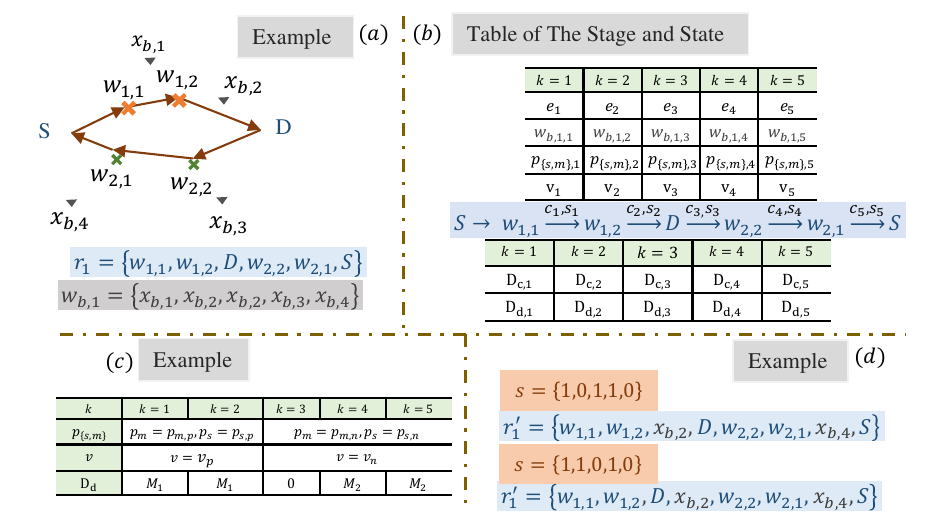}
	\caption{\textbf{(a)} The basic structure of $\mathcal{P}_1$ in the case of $N_1 = 2$ and $N_2 = 2$. \textbf{(b)} One example of the possible route of $N_1 = 2$ and $N_2 = 2$. \textbf{(c)} The table of the stage and state given the example in \textbf{(a)}. \textbf{(d)} Given the example in  \textbf{(a)}, two possible decisions of traveling to TBSs and corresponding $r_{1}^{'}$.}
	\label{fig_cost}
\end{figure*}

 The cost function of each stage is 
\begin{align}
c_k &= (D_{c,k}+D_{d,k}- D_{c,k+1})\times T_{u2b}(0) p_{s,k} +e_k(1-s_k),\nonumber
\end{align}
where $(D_{c,k}+D_{d,k}- D_{c,k+1})$ denotes the data that the UAV forwards to the TBS at stage $k$ , which is $M_{k}^{''}$, and $e_k$ denotes the additional traveling-related energy consumption,
\begin{align}
D_{c,k+1} &=  s_k(D_{c,k}+D_{d,k}),\nonumber\\
e_k &= \frac{p_{m,k}}{v_{k}}(|w_{b,i,k}-r_{i,k-1}|+|w_{b,i,k}-r_{i,k}|-|\overrightarrow{r_{i,k-1}r_{i,k}}|),
\end{align}
and our goal is to minimize the total cost,
\begin{align}
	\mathcal{P}_1:\quad c^{*} &= \min \sum_{k = 1}^{N_1+N_2+1}c_k,\label{eq_cstart}\\
	{\rm s.t.} &\quad s_{k}\in\{0,1\}, \nonumber\\
	&\quad D_{c,1} = 0, \quad D_{c,N_1+N_2+2} = 0.\label{eq_constrains}
\end{align}
in which (\ref{eq_constrains}) denotes the constraints that UAVs start without any collected data and ends with all collected data forwarded to TBSs. Besides, we also compute the additional time cost for each stage,
\begin{align}
t_k = (D_{c,k}+D_{d,k}- D_{c,k+1})\times T_{u2b}(0) +\frac{e_k(1-s_k)}{p_{m,k}}.
\end{align}
We provide an example in Fig. \ref{fig_cost}, where $N_1 = 2$ and $N_2 = 2$, to show the structure of $\mathcal{P}_1$ and how the decision modifies the route.

\begin{lemma}[Decisions of UAVs]
	Let $s^{*}$ be the solution to $\mathcal{P}_1$,
	\begin{align}
	s^{*} = \argmin \sum_{k = 1}^{N_1+N_2+1}c_k.
	\end{align}
Consequently, the overall energy consumption  and time consumption are
\begin{align}
E_{ i}^{'} &= c^{*}+E_{\rm noTBS},\nonumber\\
T_{ i}^{'} &= T_{c^{*}}+T_{\rm noTBS},
\end{align}
where $T_{ i}^{'}$ is the round trip time, $T_{c^{*}} = \sum_k t_k$ given $s^{*}$, and let $w_{b,i}^{'} = (1-s^{*}) w_{b,i}$ be the locations of TBSs that UAV travels to.
\end{lemma}
\begin{IEEEproof}
$\mathcal{P}_1$ is solved by beginning at the last stage. Let $f_n(s_n,D_{c,n}) = c_n+\sum_{i = n}^{N_1+N_2+1}c_i$ where $1\leq n\leq N_1+N_2$, and $f_{N_1+N_2+1}(s_{N_1+N_2+1},D_{c,N_1+N_2+1}) = c_{N_1+N_2+1}$, also $f_{n}^{*}(D_{c,n}) = \min_{s_n} f_n(s_n,D_{c,n})$.
Given $D_{c,n} = s_{n-1}(D_{c,n-1}+D_{d,n-1})$, we have $f_n(s_n,D_{c,n}) = c_n+f_{n+1}^{*}(s_{n},D_{c,n})$. Finally, the recursive relationship results in $f_{1}^{*}(D_{c,1}) = \min_{s_1} f_1(s_1,D_{c,1})$ where $D_{c,1} = 0$. 
	\end{IEEEproof}

If $E_{\rm i}^{'}$ is lower than $B_{\rm max}$ which means that UAVs can serve at least one more IoT clusters. If $E_{ i}^{'}$ is greater than $B_{\rm max}$, which means that UAVs are unable to transfer all the required data. Given the priority of the IoT clusters, UAVs only transfer a certain amount of data from the IoT cluster has the lowest priority,
\begin{align}
D^{'} = D_{\mathbf{w}_{\rm end}}-\frac{c^{*}+E_{\rm noTBS}-B_{\rm max}}{T_{c2u}(0)p_{s,c}+T_{u2b}(0)p_{s,d}},
\end{align}
where $p_{s,c}, p_{s,d}\in \{p_{s,n},p_{s,p}\}$  denote the serving power when collecting and delivering data, and $\mathbf{w}_{\rm end}$ denotes the collected/delivered in the given trajectory which has the lowest priority. In the case of $D^{'}<0$, it means that UAVs have enough energy to serve $N_1+N_2-1$ IoT clusters but unable to traveling to the last IoT cluster. In this case, we optimize the UAV trajectory for the first $N_1+N_2-1$ IoT clusters. In the case of $D^{'}>0$, the round trip time without optimization is 
\begin{align}
T_{\rm total,i}^{'} = T^{'}_{i}-D^{'}(T_{c2u}(0)+T_{u2b}(0)).
\end{align}
Note that $D^{'}$ is a lower bound and $T_{\rm total,i}^{'}$ is an upper bound of the given trajectory since the hovering points are not optimal. Let $M_{\rm total,i}^{'} = \sum_{i = 1}^{N_1+N_2-1}D_{c,i}+D^{'}$ denotes the collected/delivered data without optimization.

We also notice that when the UAV's energy is not enough to collect/deliver all the data and reduce  the transmission data, it actually only influences one IoT cluster, which is the one that has the lowest priority, $\mathbf{w}_{\rm end}$. Besides, if UAV increases the transmitted data for $\mathbf{w}_{\rm end}$, it only influences one TBS, which is the one UAV travels to after collecting more data from $\mathbf{w}_{\rm end}$, denoted by $w_{b,end}^{'}$. In the following text, the subscript ${end}$ denotes that this notation is used for the last IoT clusters/TBS.

Recall that we define $M_{k}^{''}$ as the delivered data for each stage and obtained by $M_{k}^{''} = (D_{c,k}+D_{d,k}- D_{c,k+1})$. Therefore, we first use Algorithm \ref{Alg_maxData} to obtain the optimal hovering points $\mathbf{h}$ given  $r_{i}^{'}$. Let $h_{b},h_{i}\in\mathbf{h}$ be the optimal points for TBSs and IoT clusters, respectively. Consequently, the additional energy is
\begin{align}
\Delta E &=  \sum_{k}p_{s,k}(1-s_{k})M_{k}^{''}(T_{u2b}(R_{u2b,k})-T_{u2b}(0))\nonumber\\
&+\sum_{l} D_{c,l}(T_{c2u}(R_{c2u,l})-T_{c2u}(0))p_{s,l},\nonumber
\end{align}
where $R_{u2b,k} = |w_{b,k}-h_{b,k}^{'}|$ denotes the horizontal transmission distance between UAVs and TBSs (since UAVs only travel to some of the TBSs in $w_b$, we simply let $h_{b,k}^{'} = w_{b,k}$ if $w_{b,k } \not\subset w_{b}^{'}$, otherwise, $h_{b,k}^{'} = h_{b,k}$), and $R_{c2u,l} = |r_{i,l}-h_{i,l}|$ denotes the horizontal transmission distance between UAVs and the $l$-th IoT clusters. 

We then maximize the transmitted data of $\mathbf{w}_{\rm end}$ while consuming all the energy of the UAV. When we optimize the transmitted data for $\mathbf{w}_{\rm end}$ and $w_{b,end}^{'}$, we ignore the influence of optimal hovering point changes of $\mathbf{w}_{\rm end}$ and $w_{b,end}$, which are $h_{i,end}$ and $h_{b,end}$, on the other hovering points $h \in \{\mathbf{h}/h_{i,end}, h_{b,end}\}$. The additional transmitted data $\Delta D$ satisfy the following equation
\begin{align}
&((M_{end}^{''}+\Delta D)T_{\rm u2b}(d_{b}^{'})-M_{end}^{''}T_{\rm u2b}(d_{b}))p_{s,d} \nonumber\\
&= ((D^{'}+\Delta D)T_{\rm c2u}(d_{i}^{'})-D^{'}T_{\rm c2u}(d_{i}))p_{s,end}, \label{eq_DData}
\end{align}
where $d_{b} = |w_{b,end}-h_{b,end}|$, $d_{i} = |\mathbf{w}_{\rm end}-h_{i,end}|$, $d_{b}^{'} = |w_{b,end}-h_{b,end}^{'}|$ and $d_{i}^{'} = |\mathbf{w}_{\rm end}-h_{i,end}^{'}|$. (\ref{eq_DData}) can be solved by using Algorithm \ref{Alg_maxData}. By doing so, we obtain the optimal hovering point $\mathbf{h}^{'}$ which is the final trajectory of the UAV given the $r_i$th trajectory.

\begin{algorithm}
	\caption{Algorithm for maximize the transmitted data}
	\SetAlgoLined
	\DontPrintSemicolon
	\KwIn{$\mathbf{w}_{\rm end},w_{b,end},h_{b,end}, h_{i,end}$: Locations}
	\KwOut{$h_{b,end}^{'}, h_{i,end}^{'}$, $\Delta D$: Locations of hovering points, additional deliver data }
	\textbf{Initialization:} $\mathbf{d}^{0}_{1}=|w_t-\mathbf{h}^{0}_1|,\mathbf{d}^{0}_{2}=|\mathbf{w}_{\rm end}-\mathbf{h}^{0}_1|,\mathbf{h}^{0} = \{h_{b,end}, h_{i,end}\}$, $r= 1$, $M_{t,1}^{0} = M_{end}^{''}$, $M_{t,2}^{0} = D^{'}$ \newline
	\SetKwFunction{FMain}{MaxData}
	\SetKwProg{Fn}{Function}{:}{}
	\Fn{\FMain{$\mathbf{w}_{\rm end},w_{b,end},h_{b,end}, h_{i,end}$}}{
		\SetKwBlock{Repeat}{repeat}{}
		The additional transmitted data is given by
		\begin{align}
		\Delta D^1 &= \frac{\Delta E}{T_{u2b}(\mathbf{d}_{1}^{0})p_{s,d}+T_{c2u}(\mathbf{d}_{2}^{0})p_{s,end}}.\nonumber
		\end{align}
		\Repeat{
			Solve (\ref{eq_Eh}) or (\ref{eq_Th}) for $\mathbf{h^{r}}$ given $\mathbf{w}_{\rm end},w_{t,d}$, the transmitted data $M_{t,1}^{r} = M_{t,1}^{r-1}+\Delta D^{r}$ and $M_{t,2}^{r} = M_{t,2}^{r-1}+\Delta D^{r}$, denote the optimal solution as $h_{t,d}^{r+1}, h_{c,end}^{r+1}$ and update $\mathbf{d^{r+1}}$\\
			Update $r = r+1$\\
			The additional energy is given by
			\begin{align}
			\Delta E &= p_{s,d}M_{t,1}^{r}(T_{\rm u2b}(\mathbf{d}_{1}^{r})-T_{\rm u2b}(\mathbf{d}_{1}^{r-1}))\nonumber\\
			&+p_{s,end}M_{t,2}^{r}(T_{\rm c2u}(\mathbf{d}_{2}^{r})-T_{\rm u2b}(\mathbf{d}_{2}^{r-1})),\nonumber\\
			\Delta D^r &= \frac{\Delta E}{T_{u2b}(\mathbf{d}_{1}^{r})p_{s,d}+T_{c2u}(\mathbf{d}_{2}^{r})p_{s,i}}+\Delta D^{r-1}.\nonumber
			\end{align}
			}
		The element-wise increase of $\mathbf{d^{r}}-\mathbf{d^{r-1}}$ is
		below a threshold $\epsilon>0$\;
	}\textbf{End Function}
	\label{Alg_maxData}
\end{algorithm}

In this way, we obtain the optimal trajectory of a given route $r_i$, which maximizes the transmitted data/minimizes the round trip time within limited UAV battery. Recall that $T_{\rm total,i}$ and $M_{\rm total,i}$ are the minimal time and maximal data of the $r_i$the trajectory,
\begin{align}
M_{\rm total,i} &= M_{\rm total,i}^{'}+\Delta D,\nonumber\\
T_{\rm total,i} &= T_{\rm total,i}^{'}+\Delta T,\nonumber
\end{align}
where,
\begin{align}
	\Delta T = \Delta D(T_{c2u}(d_{i}^{'})+T_{u2b}(d_{b}^{'})-T_{c2u}(d_{i})-T_{u2b}(d_{b})).
\end{align}
Consequently, the minimal time path and maximal data path of a give realization are obtained by selecting from all the routes $\mathbf{r}$, as given in Definition \ref{Def_MinTime}.
%
\subsection{Optimal Trajectory}
In this part, we propose the final algorithms which integrated all the previous algorithms and finalize the UAV trajectory from the first step. We solve the minimal time path and maximal data path separately, e.g., when we looking for the maximal data path, we solve the optimization problems based on (\ref{eq_Eh}).

\begin{algorithm}
	\caption{Algorithm for UAV trajectory}
	\SetAlgoLined
	\DontPrintSemicolon
	\KwIn{$N_1$, $N_2$: Number of serving IoT clusters\\
	      $x_1$, $x_2$, $x_t$, $S$, $D$: Locations }
	\KwOut{$T_{\rm total},M_{\rm total},\mathbf{h}$ }
	\textbf{Initialization:} $w =1$ \newline
	\SetKwFunction{FMain}{UAVTraj}
	\SetKwBlock{Repeat}{repeat}{}
	\SetKwProg{Fn}{Function}{:}{}
	\Fn{\FMain{$N_1$, $N_2$, $x_1$, $x_2$, $x_t$, $S$, $D$}}{
		Using Algorithm \ref{Alg_iotlocation} to obtain $\mathbf{w}$, the locations and priorities of IoT clusters\\
		\Repeat{
		$\mathbf{w^{'}} = \mathbf{w}(1:w)$ and $\mathbf{r}$ contains all the permutations of $\mathbf{w^{'}}$\\
		\ForEach{$ i \leq (w+1)!$}{
			Solve (\ref{eq_wt}) for $r_i$ to obtain $w_{b,i}$\\
			Solve $\mathcal{P}_1$ to obtain the decision of traveling to TBSs, $T_{i}^{'}$ and $E_{i}^{'}$\\
		}
	    Find the minimal time path or minimal energy path, the time and energy consumption of the path are $T^{'}$ and $E^{'}$, respectively\\
		{\eIf{$ E^{'} \leq B_{\rm max} $}
			{Update $w = w+1$}
			{Use Algorithm \ref{Alg_path} to obtain the hovering point for collecting/delivering data\\
			Use Algorithm \ref{Alg_maxData} to obtain the maximum transmitted data and $E_{\rm total}=B_{\rm max}$} }
		}
	$E_{\rm total}=B_{\rm max}$\
	}\textbf{End Function}
	\label{Alg_overall}
\end{algorithm}

 In Algorithm \ref{Alg_overall}, we find a minimal energy path\footnote{Note that the term "minimal energy path" refers to the trajectory followed by the UAV  that minimizes energy consumption during travel/hovering, with or without a package. The energy saved through this efficient path is then allocated for data transmission. Consequently, the minimal energy path leads to an optimized data transmission route, enabling maximum data throughput.} since minimal energy results in maximal transmitted data (since UAV saves energy from traveling, hence, it has more energy on data transmission), and we only optimize the hovering points for the path with minimal time or minimal energy (given the optimization objective in Definition \ref{Def_MinTime}) since the differences are negligible compared with the differences between different routes.  	Note that the predefined route (path with minimal time or minimal energy) is found by a greedy-based method, even though the visiting sequence and hovering points also influence energy consumption. The reason for separating is that optimizing hovering points has a very limited influence on the energy consumption compared to the different visiting sequences and we will show this in the numerical results section. 

The above Algorithm \ref{Alg_overall} conditions on the realizations of locations of IoT clusters and TBSs, and we are interested in the average performance of the system and data transmission efficiency, minimum round trip time in general case. However, it is difficult to obtain the joint PDF of the locations. Instead, we apply Monte-Carlo simulations with a large number of iterations to obtain the average performance.

\section{Numerical Results}

In this section, we validate our analytical results with simulations and evaluate the impact of various system parameters on the network performance. Unless stated otherwise, we use the simulation parameters as listed herein Table \ref{par_val}.
\begin{table}[ht]\caption{Table of Parameters}\label{par_val}
	\centering
	\begin{center}
		\resizebox{1\columnwidth}{!}{
			\renewcommand{\arraystretch}{1}
			\begin{tabular}{ {c} | {c} | {c}  }
				\hline
				\hline
				\textbf{Parameter} & \textbf{Symbol} & \textbf{Simulation Value}  \\ \hline
				Density of TBSs and type-I $\&$ II IoT clusters & $\lambda_{b}$, $\lambda_{i,1}$ $\&$  $\lambda_{i,2}$ & 1, 1 $\&$ 5 km$^{-2}$ \\ \hline
				Average package weight & $\bar{w}$ & $1$ kg \\ \hline
				Required data of type-I $\&$ II IoT clusters & $M_1$ $\&$ $M_2$ & $2200$ $\&$  $600$ bit/Hz\\ \hline
				IoT cluster radius & $r_c$ & 50  m\\ \hline
				Optimal with/without package velocity & $v_p,v_n$ & 20, 18 m/s\\ \hline
				Serving-related power (with/without package) & $p_{s,p},p_{s,n}$ & 252 J, 178 J\\ \hline
				Traveling-related power (with/without package) & $p_{m,p},p_{m,n}$ & 193 J, 159 J \\ \hline
				UAV altitude & $h_u$ & 100 m\\\hline
				Battery capacity & $B_{\rm max}$ & 177.6 W$\cdot$H \\\hline
				N/LoS environment variable & $a, b$ & 4.9 0.43 \\\hline
				Transmission power & $\rho_{i},\rho_{u}$ & 0.1 mW, 0.1 W\\\hline
				Noise power & $\sigma^2 $ & $10^{-9}$ W\\\hline
				N/LoS path-loss exponent & $\alpha_{n},\alpha_{ l}$ & $4,2.1$ \\\hline
				N/LoS fading gain & $m_{ n},m_{ l}$ & $1,3$ \\\hline
				N/LoS additional loss& $\eta_{ n},\eta_{l}$ & $-20,0$ dB 
				\\\hline\hline
		\end{tabular}}
	\end{center}
\end{table}
\begin{figure*}
	\centering
	\subfigure[]{\includegraphics[width = 0.8\columnwidth]{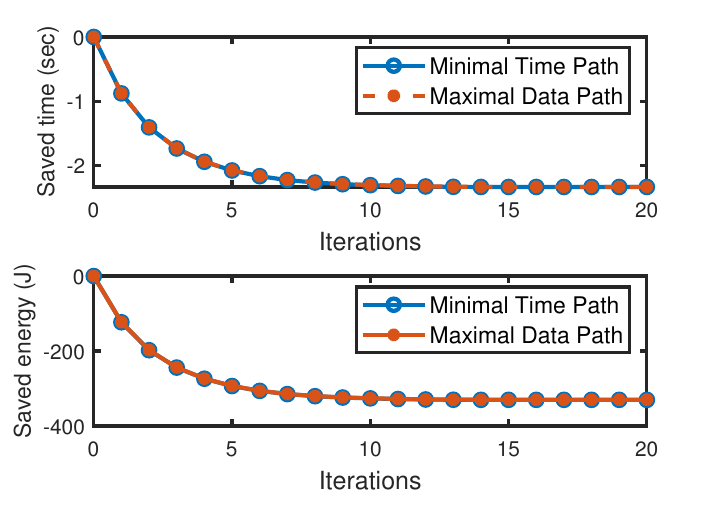}}
	\subfigure[]{\includegraphics[width = 0.8\columnwidth]{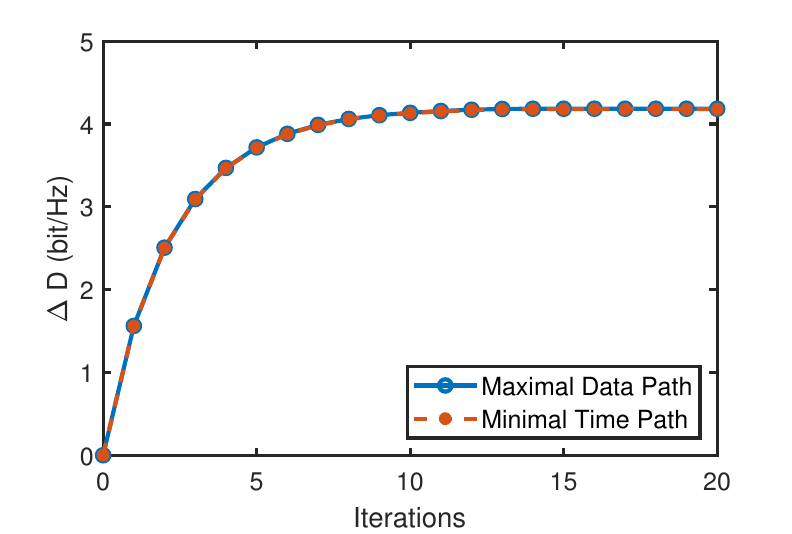}}
	\caption{Convergence behaviour of the last step of the proposed Algorithm \ref{Alg_overall}.}
	\label{fig_iterations}
\end{figure*}

	In Fig. \ref{fig_iterations}, we plot the convergence behavior of the last step and optimize the hovering points, of the algorithm. As discussed in the previous paragraph, separating the process of finding the optimal hovering points from the process of finding the predefined route has limited influence on the final results, which is shown in Fig. \ref{fig_iterations} (b), while for the latter step, different visiting sequences always results in a different number of served IoT clusters. 

\begin{figure*}
	\centering
	\subfigure[]{\includegraphics[width = 1.6\columnwidth]{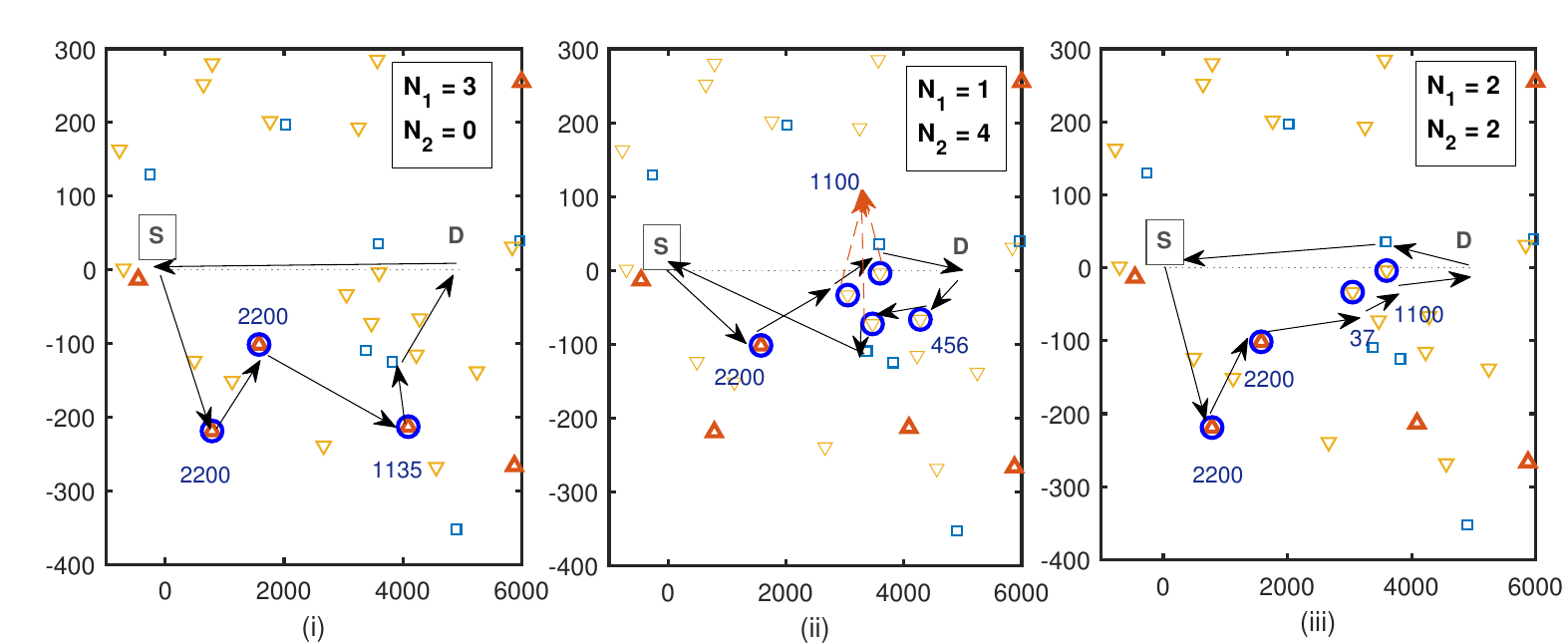}}
	\subfigure[]{\includegraphics[width = 1.6\columnwidth]{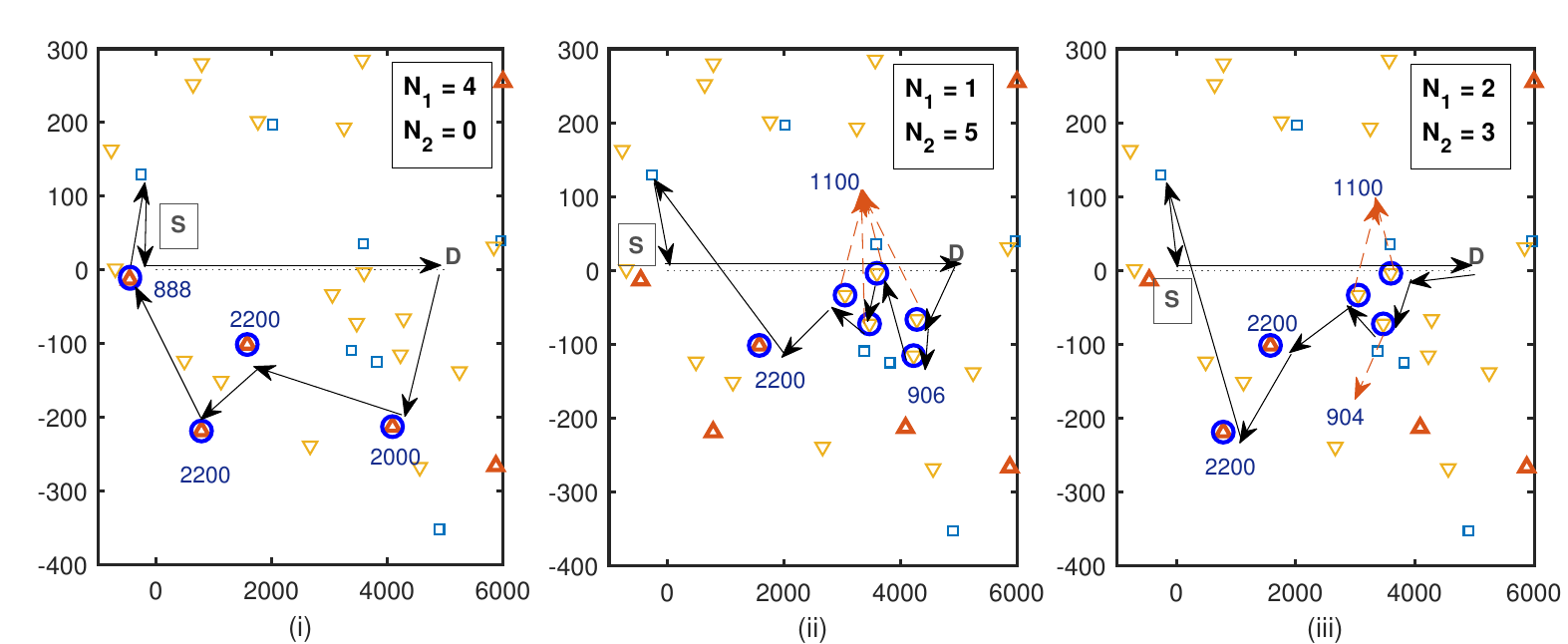}}
	\caption{Realizations of the optimal trajectory of UAV, \textbf{(a)} minimal time path and, \textbf{(b)} maximal data path. Solid arrows show the trajectory of UAVs which start at ${\rm S}$, square markers denote TBSs and two types of triangular markers denote two types of IoT clusters, in which up-pointing triangular denote type-I IoT clusters which have higher priority and more data to be collected than down-pointing triangular. Circle markers denote the IoT clusters that UAVs served and the number near each circle markers are the collected data.}
	\label{fig_path}
\end{figure*}

For the simulation of the considered system setup, we first generate three independent PPP realizations to model the locations of the TBSs and two types of IoT cluster centers. To compute the transmission time between UAVs and IoT clusters, and TBSs, we generate the locations of IoT devices that are uniformly distributed within the IoT clusters. For each device, we compute its SNR and transmission time, and the total time of the cluster is obtained by summing up the individual transmission time. We then compute the round trip time and collected/delivered data size for both minimal time path and maximal data path given the realizations. Finally, we apply Monte-Carlo simulations with a large number of iterations to ensure accuracy.

Before we analyze the collected/delivered data and needed time, we show the convergence behaviour of the last step of the proposed Algorithm \ref{Alg_overall}. Since  Algorithm \ref{Alg_overall} integrated all the algorithms in this work, we only need to show the convergence of Algorithm \ref{Alg_overall}, which is about the optimization of the hovering points. For a given realization if $\Phi_{i,1},\Phi_{i,2}$ and $\Phi_b$, we can observe that the additional transferred data/saved round trip time increases/decreases quickly with the number of iterations and the algorithm converges at about $15$ iterations.

In Fig. \ref{fig_path}, we provide a numerical example to demonstrate the effectiveness of the proposed algorithm. For a random realization of locations, UAVs serve the IoT clusters based on the predefined $N_1$, and use the remaining energy to serve $N_2$ type-II IoT clusters. 
 The values of $N_1$ and $N_2$ are not predetermined or specific in nature. Instead, they serve as parameters that can be adjusted based on the characteristics of the particular IoT network under consideration. These values are open to adaptation and customization to suit different scenarios, ensuring the method's versatility in addressing various network configurations. 

Fig. \ref{fig_path} $(a)$ shows the minimal time path, where UAVs' trajectory is the shortest path among all the possible paths. Fig. \ref{fig_path} $(b)$ shows the maximal data path, where UAVs prefer to deliver the package first and then collect/deliver the data because of the higher energy consumption of UAVs with the package. It is worth mentioning that among all the optimal trajectories in different realizations, we observe that UAVs prefer to forward all the collected data together to one TBS, say $w_{b,end}^{'}$, which is the closest to the route that the UAV finish collecting all the data and either traveling to deliver the package or back to ${\rm S}$. The reason is since UAVs must deliver all the collected data, traveling to $w_{b,end}^{'}$ is mandatory. Therefore, it is inefficient to travel to multiple TBSs while the overall transmission time (between the UAV and TBSs) is approximately the same (Because the transmission time is also a function of the horizontal transmission distance, but it does not cause a huge gap. As shown in Fig. \ref{fig_iterations}, the time consumption difference between the optimal and initial points is relatively small.) but increasing the travel-related time and energy consumption.

\begin{figure}[ht]
	\centering
	\subfigure[]{\includegraphics[width = 1\columnwidth]{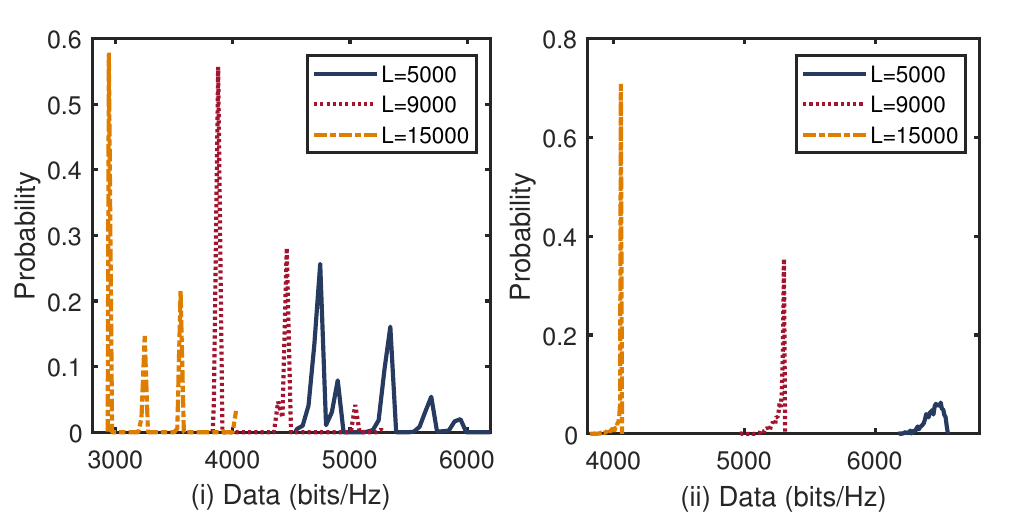}}
	\subfigure[]{\includegraphics[width = 1\columnwidth]{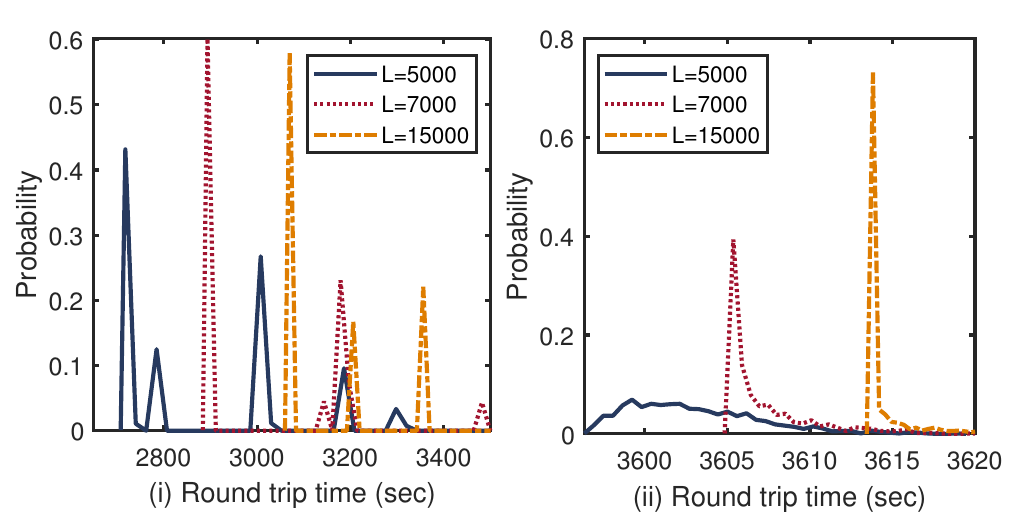}}
	\caption{\textbf{(a)} PDF of the collected/delivered data under different ${\rm S-D}$ distances in the case of \textbf{(a-i)} minimal time path, and \textbf{(a-ii)} maximal data path. \textbf{(b)}  PDF of the round trip time under different $S-D$ distances in the case of \textbf{(b-i)} minimal time path, and \textbf{(b-ii)} maximal data path.}
	\label{fig_datapdf}
\end{figure}
In Fig. \ref{fig_datapdf}, we plot PDF of the collected/delivered data and round trip time under different ${\rm S-D}$ distances. As expected, for both optimal UAV trajectories, collected/delivered data decreases, while the round trip time increases with the increase of the ${\rm S-D}$ distances. By comparing the minimal time path and maximal data path, we observe that the time consumption of the minimal data path is significantly lower than the maximal data path. Interestingly, we observe that the round trip time and collected/delivered data of the minimal time path have several peaks while the maximum data path does not. This is because of the UAV velocity. As mentioned in Section \ref{sec_power}, the optimal velocities with/without the package that minimizes the energy consumption are different. For a minimal time path, UAVs serve the IoT clusters based on the shortest trajectory. It means that UAVs can deliver the package at any part of the trajectory. While the overall traveling distances are almost the same, the velocity changes after delivering the package. However, for maximal time paths, UAVs always deliver the package first to save energy, and the round trip time is continuous without gaps.


\begin{figure}[ht]
	\centering
    \subfigure[]{\includegraphics[width = 0.8\columnwidth]{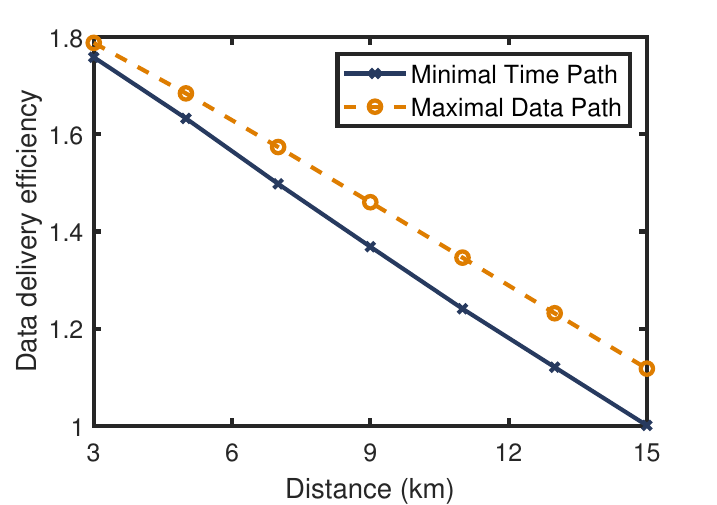}}
     \subfigure[]{\includegraphics[width = 0.8\columnwidth]{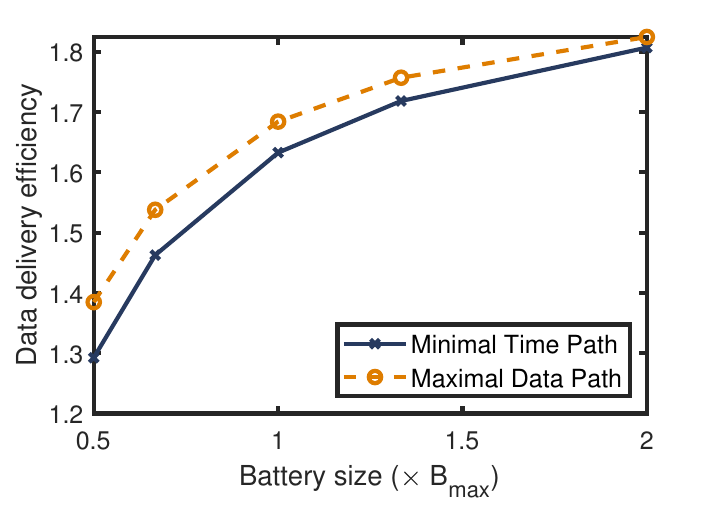}}
	\caption{Data transmission efficiency under, \textbf{(a)} different ${\rm S-D}$ distances and,  \textbf{(b)} different battery size, when $L = 5$ km.}
	\label{fig_dataeff}
\end{figure}
It is difficult to tell from Fig. \ref{fig_datapdf}   which optimal trajectory is better. To further investigate the efficiency of data delivery, we plot the collected/delivered data over the round trip time, as defined in Definition \ref{Def_dataeff}, data delivery efficiency, in Fig. \ref{fig_dataeff}. Obviously, the maximal data path has higher efficiency. With the increase of ${\rm S-D}$ distance, data delivery efficiencies decrease for both paths, which is because UAVs consume more energy in package delivery. With the increase of the battery size, data delivery efficiency increases rapidly at first and then slow down since it reaches the system's upper bound, which is limited by the SNR.. While maximal data path achieves higher data delivery efficiency, minimal time path enables UAVs to finish a round trip within a short time, therefore, delivering more packages.

 Finally, to illustrate the benefit of integrating multiple purposes into a single UAV, we compare the multi-purpose drone with a UAV that can only finish one task at a time, e.g., a single-purpose UAV needs two trips, one for package delivery and one for data delivery. The time, energy consumption, and delivery efficiency are compared in Table \ref{table_comp} under $L = 5$ km, and the `time' and `data' in the bracket denote the minimal time path and maximal data path, respectively. By integrating multiple purposes into a single UAV, the multi-purpose drone streamlines operations, saves time, reduces energy consumption, and enhances overall delivery efficiency. These advantages make it a more effective and economical choice compared to single-purpose UAVs that can only handle one task at a time.
	\begin{table}\caption{Comparison between multi-purpose and single-purpose drone}\label{table_comp}
	\vspace{-8mm}
	\centering
	\begin{center}
		\resizebox{1\columnwidth}{!}{
			\renewcommand{\arraystretch}{1}
			\begin{tabular}{ {c} | {c}  }
				\hline
				\hline
				\textbf{Method} & \textbf{Time, Energy Consumption} \\ \hline
				Multi-purpose (data) & (3.6$\times10^3$ s, 6.4$\times10^5$ J,)\\ \hline
				Multi-purpose (time) & (2.9$\times10^3$ s, 6.4$\times10^5$ J,)\\ \hline
				Single-purpose (data)  & (4.2$\times10^3$ s, 7.6$\times10^5$ J)\\ \hline
				Single-purpose (time)  & (3.9$\times10^3$ s, 7.6$\times10^5$ J)
				\\\hline
				\textbf{Method} &  \textbf{Time, Delivered Data, Delivery Efficiency}\\ \hline
				Multi-purpose (data) & (3.6$\times10^3$ s, 6.1$\times10^3$ bit/Hz, 1.69)\\ \hline
				Multi-purpose (time) & (2.9$\times10^3$ s, 4.8$\times10^3$ bit/Hz, 1.65) \\ \hline
		    	Single-purpose (data)  & (4.2$\times10^3$ s, 6.7$\times10^3$ bit/Hz, 1.59)\\ \hline
		    	Single-purpose (time)  & (3.9$\times10^3$ s, 6.2$\times10^3$ bit/Hz, 1.58)
				\\\hline
				\hline
		\end{tabular}}
	\end{center}
\end{table} 
\section{Conclusion}

This paper presented a novel system model in which UAVs simultaneously serve for multiple tasks: data collection/delivery and package delivery. We used tools from stochastic geometry and optimization to investigate the feasibility of UAVs functioning as wireless data relays and means of transportation at the same time. Specifically, we considered UAVs to collect/deliver data for multiple IoT clusters while delivering packages. We proposed some algorithms which enable UAVs to select the served IoT clusters based on the distance and priority. Moreover, the proposed algorithms allow UAVs to predefine the trajectory to either minimize the round trip time or maximize the collected/delivered data while serving multiple IoT clusters.  

This work tapped a new aspect of the applications of UAVs, integrated multi-function on one UAV, and we proposed algorithms which enable UAV serving multiple IoT clusters. Instead of dedicated UAVs, multipurpose drones seem more efficient and realistic in real life, which can highly reduce the aerial traffic and conflicts in future networks. While UAVs are widely used in last-mile deliveries, they can also be used in wireless communication networks to fully display their benefits: flexibility, capability to optimize their locations in real-time, and additional cellular network capacity.

\bibliographystyle{IEEEtran}
\bibliography{Ref11}

\end{document}

%% file: notation.tex
\def\nb0{{\mathbf{0}}}
\def\nb1{{\mathbf{1}}}

\newtheorem{lemma}{Lemma}

\newtheorem{definition}{Definition}

\newtheorem{theorem}{Theorem}

\def\argmin{\operatorname{arg~min}}

\def\R{\mathbb{R}}

